\documentclass[nobibnotes,twocolumn,superscriptaddress,longbibliography,nobalancelastpage]{revtex4-2}

\usepackage{amsmath}
\usepackage{amssymb}
\usepackage{graphicx}
\usepackage{braket}
\usepackage{hyperref}
\newcommand{\down}{\downarrow}
\newcommand{\up}{\uparrow}

\usepackage{xr}
\begin{document}

\title{
Observation of a topological defect state in the quantum Rabi model
}

\author{Kyungmin Lee}
\thanks{These authors contributed equally to this work.}
\affiliation{\mbox{Department of Computer Science and Engineering, Seoul National University, Seoul 08826, Republic of Korea}}
\affiliation{\mbox{Automation and System Research Institute, Seoul National University, Seoul 08826, Republic of Korea}}
\affiliation{\mbox{NextQuantum, Seoul National University, Seoul 08826, Republic of Korea}}

\author{Jiyong Kang}
\thanks{These authors contributed equally to this work.}
\affiliation{\mbox{Department of Computer Science and Engineering, Seoul National University, Seoul 08826, Republic of Korea}}
\affiliation{\mbox{Automation and System Research Institute, Seoul National University, Seoul 08826, Republic of Korea}}
\affiliation{\mbox{NextQuantum, Seoul National University, Seoul 08826, Republic of Korea}}

\author{Sunkyu Yu}
\affiliation{Intelligent Wave Systems Laboratory, Department of Electrical and Computer Engineering, Seoul National University, Seoul 08826, Republic of Korea}

\author{Jaehun You}
\affiliation{\mbox{Department of Computer Science and Engineering, Seoul National University, Seoul 08826, Republic of Korea}}
\affiliation{\mbox{Automation and System Research Institute, Seoul National University, Seoul 08826, Republic of Korea}}
\affiliation{\mbox{NextQuantum, Seoul National University, Seoul 08826, Republic of Korea}}

\author{Wonhyeong Choi}
\affiliation{\mbox{Department of Computer Science and Engineering, Seoul National University, Seoul 08826, Republic of Korea}}
\affiliation{\mbox{Automation and System Research Institute, Seoul National University, Seoul 08826, Republic of Korea}}
\affiliation{\mbox{NextQuantum, Seoul National University, Seoul 08826, Republic of Korea}}

\author{Taehyun Kim}
\email[Corresponding author: ]{taehyun@snu.ac.kr}
\affiliation{\mbox{Department of Computer Science and Engineering, Seoul National University, Seoul 08826, Republic of Korea}}
\affiliation{\mbox{Automation and System Research Institute, Seoul National University, Seoul 08826, Republic of Korea}}
\affiliation{\mbox{NextQuantum, Seoul National University, Seoul 08826, Republic of Korea}}
\affiliation{\mbox{Institute of Applied Physics, Seoul National University, Seoul 08826, Republic of Korea}}
\affiliation{\mbox{Inter-university Semiconductor Research Center, Seoul National University, Seoul 08826, Republic of Korea}}
\affiliation{\mbox{Institute of Computer Technology, Seoul National University, Seoul 08826, Republic of Korea}}

\begin{abstract}
Synthetic lattices built from hybrid qubit--oscillator systems provide a platform for exploring topology, with multiple physical degrees of freedom enabling control over lattice geometry.
A single spin--oscillator pair, described by the quantum Rabi model, provides a minimal system supporting a topological defect state on a lattice formed from the oscillator's infinite ladder of number states.
However, the realization of the defect state and its robustness remain experimentally unexplored.
Here we realize a topological defect state of the quantum Rabi model in a single trapped $^{171}\mathrm{Yb}^{+}$ ion.
As a coupling phase varies, reconstructed phase-space distributions yield centroid trajectories with windings of one and zero in two regimes, linking discrete-lattice topology to the geometry of the oscillator's continuous position--momentum space.
The measured spin response to drive-amplitude modulation relates the stability of spin polarization to energy gaps and allowed transitions.
Together, these results establish a minimal quantum system for realizing and probing topological defect states.
\end{abstract}
\maketitle

\section{Introduction}
\label{sec:introduction}

\begin{figure*}[t!]
\centerline{\includegraphics[width=1.\linewidth]{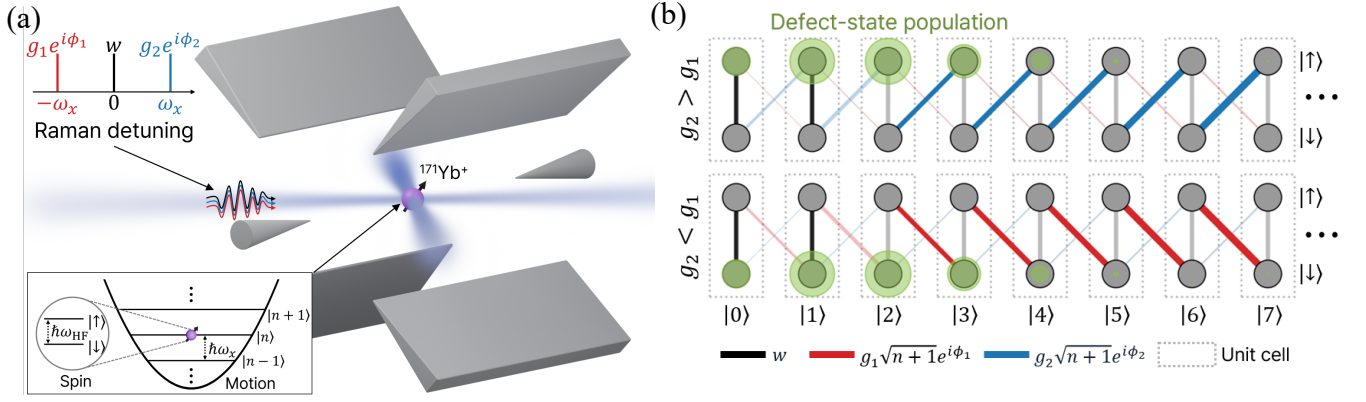}}
\caption{
Quantum Rabi model in a trapped-ion Fock-state lattice.
(a) A three-tone Raman drive couples the internal spin and one motional mode of a single $^{171}\mathrm{Yb}^{+}$ ion.
The carrier sets $w$, while the red- and blue-sideband tones set $g_1e^{i\phi_1}$ and $g_2e^{i\phi_2}$ in Eq.~(\ref{eqn:eqn1}).
(b) Spin--Fock-state lattice generated by these couplings.
The upper and lower rows show the regimes $g_2>g_1$ and $g_1>g_2$, respectively.
Black vertical links denote the carrier coupling. Red and blue diagonal links denote the two sideband couplings, whose strengths increase with Fock number.
Green markers indicate the calculated defect-state population on the $\ket{\uparrow}$ or $\ket{\downarrow}$ spin sublattice in the respective regimes.
The populations are calculated for $w/g_{\max}=1.6$, $g_{\min}/g_{\max}=0.25$, and $\phi_1=\phi_2=0$, where $g_{\max}=\max(g_1,g_2)$ and $g_{\min}=\min(g_1,g_2)$.
}
\label{fig:figure1}
\end{figure*}

Topological phases, which are characterized by global invariants and robust boundary states, have been explored across various quantum platforms~\cite{rev_topology3,ssh_ion,ssh_photon_rev,ssh_neutral_atom_rev}.
Hybrid qubit--oscillator systems also offer a versatile platform for exploring topological degrees of freedom by forming synthetic lattices, known as Fock-state lattices (FSLs), through spin--boson interactions: oscillator number states act as lattice sites, and the interactions induce site-dependent hopping~\cite{hybrid_qc_device,hybrid_cavity_qed,hybrid_circuit,hybrid_iontrap,fsl_scully,fsl_cai_and_wang,fsl_saugmann_and_larson,fsl_xing,fsl_driven_gauge_transitions,fsl_thouless_squeezed}.
Experiments on multimode FSLs have demonstrated the preparation, reconstruction, and transport of topological states~\cite{fsl_quantum_topology_light,fsl_photonic_pumping}, while coupled quantum devices have realized multidimensional FSLs~\cite{fsl_ab_cages}.
Multiple internal levels of a trapped ion have also enabled synthetic lattices with tunable artificial gauge fields~\cite{ion_nonabelian_caging}.

Although these additional degrees of freedom enable a variety of synthetic lattice geometries, a single bosonic mode already provides an infinite ladder of number states that can serve as lattice sites.
Interaction with a single spin, as described by the quantum Rabi model (QRM), introduces coupling between these states, forming an FSL in a minimal quantum system~\cite{QRM_Review,generalized_QRM,fsl_cai_and_wang,fsl_saugmann_and_larson}.
The spin--boson couplings that form this FSL have been implemented in experiments on dissipative preparation and coherent control of bosonic states~\cite{kienzler_reservoir_engineering,kienzler_squeezed_fock,bosonic_logical_states}, as well as chiral transport induced by artificial magnetic fields~\cite{ion_synthetic_dimension_flux}.
Using a single spin and a single bosonic mode, quantum-walk experiments have also simulated bulk band topology and measured the associated Berry phases~\cite{flurin_phase_space_topology}.
The QRM can also exhibit intrinsic topological structure, manifested in a localized defect state at an interface formed by competition between a spin drive and the oscillator's two number-dependent couplings.
The defect state occupies opposite spin sublattices in the two regimes distinguished by which spin--boson interaction is stronger~\cite{fsl_lee}.
However, the realization of the defect state in both coupling regimes and its robustness under controlled perturbations remain experimentally unexplored.
This requires an experimental approach that combines preparation of the defect state with complementary measurements of the bosonic mode and spin to identify its topological character and probe its robustness, respectively.

Here we experimentally demonstrate a topological defect state of the resonantly driven QRM in a single trapped $^{171}\mathrm{Yb}^{+}$ ion, using a single spin and a single motional mode.
We prepare the defect state in both regimes distinguished by the relative strengths of the two spin--boson couplings.
Quantum-state tomography yields bosonic phase-space distributions consistent with those of the target defect state.
As a coupling phase is varied, the centroids of these distributions trace trajectories with winding numbers of one and zero in the two regimes, providing a phase-space readout of the topological structure.
We further probe the stability of the defect state's spin polarization under controlled modulation of the drive amplitude, observing a suppressed response away from excitation resonances.
Our work links topological structure in a discrete Fock-state lattice to the measurable geometry of the oscillator's continuous position–momentum phase space, and relates the stability of spin polarization to the system's energy gaps and allowed transitions.
Together, these results establish a single spin coupled to a single bosonic mode as a minimal quantum system for realizing and probing topological defect states.

\section{Results}

We consider a system described by~\cite{generalized_QRM,fsl_lee}
\begin{align}
\hat H/\hbar={}&w\hat\sigma_x+\left[g_1e^{i\phi_1}\hat a^\dagger\hat\sigma_-
+g_2e^{i\phi_2}\hat a\hat\sigma_-+\mathrm{H.c.}\right],
\label{eqn:eqn1}
\end{align}
where $\hat a$ annihilates a motional quantum and $\hat\sigma_-=\ket{\down}\bra{\up}$.
We take $w,g_1,g_2$ to be real and positive and use the carrier phase as the spin reference.
The Hamiltonian in Eq.~\eqref{eqn:eqn1} is the driven anisotropic QRM with vanishing effective oscillator and spin on-site energies.
Resonant Raman frequencies implement this limit in a common rotating frame; Appendix~\ref{Appendix_ion_trap_system} explains how the resonant
drive conditions eliminate the effective on-site terms.

The hyperfine states $\ket{\down}=\ket{F=0,m_F=0}$ and $\ket{\up}=\ket{F=1,m_F=0}$ encode the spin, and a radial mode at $\omega_x=(2\pi)~910~\mathrm{kHz}$ supplies the Fock basis.
Three phase-coherent Raman tones drive the carrier and the two first sidebands, as shown in Fig.~\ref{fig:figure1}(a), providing independent amplitude and phase control of the couplings in Eq.~\eqref{eqn:eqn1}.
Details of the experimental setup are provided in Appendix~\ref{Appendix_ion_trap_system}.

In the FSL, the carrier connects the two spin states at fixed $n$, while the sidebands connect neighboring Fock levels with hopping amplitudes proportional to $\sqrt{n+1}$, as shown in Fig.~\ref{fig:figure1}(b).
Their strength relative to $w$ therefore changes along the lattice, creating a tunable SSH-like interface that supports a normalizable zero-energy defect state for $g_1\ne g_2$~\cite{fsl_lee}.
The state occupies $\ket{\down}$ for $g_1>g_2$ and $\ket{\up}$ for $g_1<g_2$, as illustrated by the calculated populations in Fig.~\ref{fig:figure1}(b).

\subsection{Experimental realization}
\label{sec:adiabatic_prepare}

The target defect state has the form
\begin{equation}
\ket{\psi_0}=\hat U(\theta,\alpha,r)\ket0\ket A,
\quad \hat U=\hat R(\theta)\hat D(\alpha)\hat S(r),
\label{eqn:eqn2}
\end{equation}
where the occupied sublattice can be identified by $\ket{A} \in \{\ket{\up},\ket{\down}\}$.
The motional component is a rotated, displaced, and squeezed vacuum; the operators and their drive-dependent parameters are defined in Appendix~\ref{Appendix_adiabatic_preparation}.

After Doppler and sideband cooling, we prepare $\ket{0}\ket{A}$ using spin initialization and microwave control~\cite{RMP_iontrap}.
We switch on the stronger sideband, for which this state is dark, and ramp the carrier and weaker sideband to $w=r_c g_{\max}$ and $g_{\min}=r_s g_{\max}$.
Here, $g_{\max}=\max(g_1,g_2)$, $g_{\min}=\min(g_1,g_2)$, and $r_c$ and $r_s$ specify the target coupling ratios.
We use $g_{\max}/(2\pi)=4.5~\mathrm{kHz}$ and $r_s=0.25$.
A sinusoidal ramp of duration $0.5~\mathrm{ms}$ prepares the states for characteristic function tomography, while a $1~\mathrm{ms}$ ramp precedes the modulation measurements.
Simulation results regarding the adiabatic preparation process are presented in Appendix~\ref{Appendix_adiabatic_preparation}.
The sideband phases are calibrated independently using a spin-dependent displacement sequence~\cite{ion_sdf}; see Supplementary Fig.~\ref*{supp-fig:figure_calibration}.

\begin{figure}[b!]
\centering
\includegraphics[width=\columnwidth]{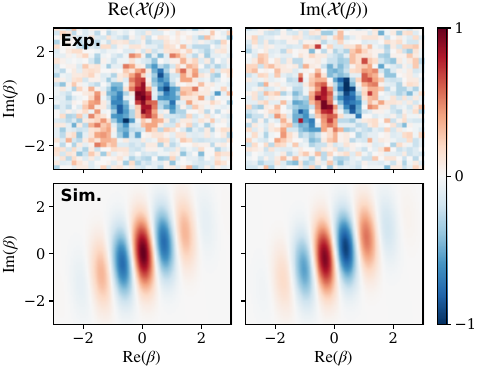}
\caption{Characteristic-function tomography for $g_1>g_2$, $r_c=2$, $r_s=0.25$, $\phi_1=-\pi/3$, and $\phi_2=0$. Columns show the real and imaginary parts of $\mathcal X(\beta)$. Rows show experiment (Exp.) and the ideal-state calculation (Sim.) from Eq.~\eqref{eqn:eqn2}. The conditional motional-state fidelity inferred from their overlap is $0.915\pm0.014$, where the uncertainty denotes one standard deviation.}
\label{fig:figure2}
\end{figure}

After postselecting the occupied spin sublattice $\ket A$, we reconstruct the motional characteristic function $\mathcal X(\beta)=\operatorname{Tr}[\rho_{\mathrm{mot}}\hat D(\beta)]$ using characteristic-function tomography~\cite{tomography_ion}, where $\rho_{\mathrm{mot}}$ is the normalized conditional motional state.
Figure~\ref{fig:figure2} compares a representative measured characteristic function with that of the motional component of the target defect state in Eq.~\eqref{eqn:eqn2}.
Their overlap gives a conditional motional-state fidelity of $F=0.915\pm0.014$ (one standard deviation).
The corresponding measurements for both coupling regimes and all phase settings used in Fig.~\ref{fig:figure3} are given in Supplementary Figs.~\ref*{supp-fig:figure_char_rc1_first}--%
\ref*{supp-fig:figure_char_rc2_second}.

\subsection{Probing phase-space winding}
\label{sec:winding_measure}

The Wigner-function centroid of the defect state is~\cite{fsl_lee}
\begin{equation}
\gamma_c=\langle\hat a\rangle=
\frac{w(g_2e^{-i\phi_2}-g_1e^{i\phi_1})}{g_1^2-g_2^2}.
\label{eqn:eqn3}
\end{equation}
At fixed $\phi_2=0$, varying $\phi_1$ through $2\pi$ traces a circle with radius $R=|wg_1/(g_1^2-g_2^2)|$ and center at a distance $C=|wg_2/(g_1^2-g_2^2)|$ from the origin.
Thus $R>C$ for $g_1>g_2$, while $R<C$ for $g_1<g_2$.
The winding \(W=\frac{1}{2\pi}\oint d\arg\gamma_c\), with the integral taken along the corresponding centroid trajectory, is consequently one or zero, respectively.
For the ideal defect-state family and the phase sweep defined above, these windings satisfy \(W=(1-z_A)/2\), where \(z_A=\langle A|\hat{\sigma}_z|A\rangle\) and $\hat\sigma_z=\ket{\up}\bra{\up}-\ket{\down}\bra{\down}$.
Thus \(W\) distinguishes the occupied spin sublattice in the two coupling regimes, both of which support a normalizable defect state~\cite{fsl_lee}.
Changing the positive carrier amplitude $w$ scales the trajectory without changing its winding.

\begin{figure}[t!]
\centering
\includegraphics[width=\columnwidth]{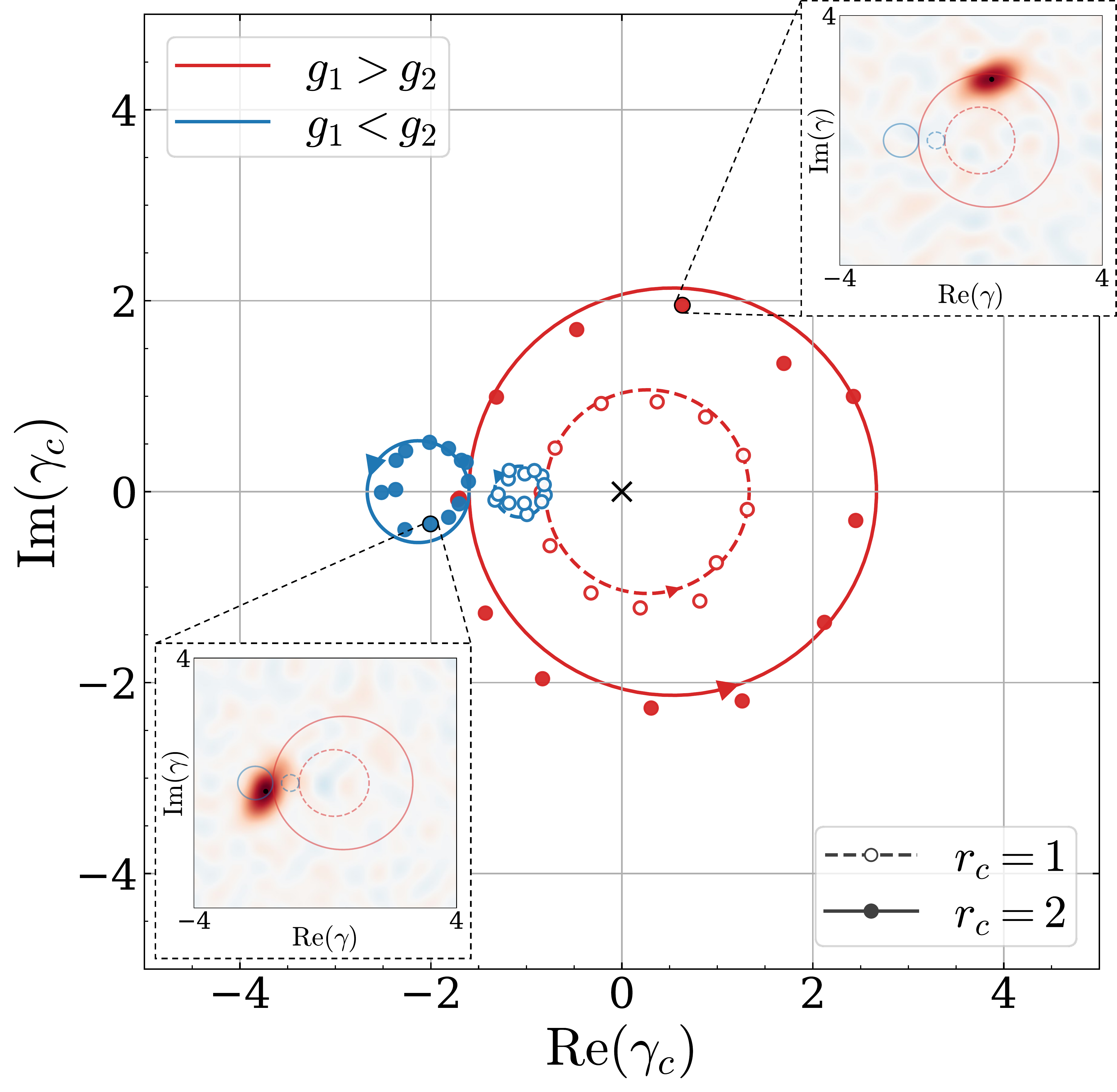}
\caption{Measured phase-space winding for $\phi_2=0$ and $r_s=0.25$. Open and filled circles show fitted Wigner centroids for $r_c=1$ and $2$. Dashed and solid curves are the corresponding predictions of Eq.~\eqref{eqn:eqn3}. Red trajectories ($g_1>g_2$) enclose the origin once, while blue trajectories ($g_1<g_2$) do not. Arrows indicate increasing $\phi_1$. Insets show selected reconstructed Wigner functions with fitted centroids marked by black dots and the predicted trajectories overlaid.}
\label{fig:figure3}
\end{figure}

To map these trajectories, we independently prepare the defect state at each sampled value of $\phi_1$, with $\phi_2=0$, and measure its characteristic function.
We then Fourier transform the measured characteristic functions to obtain Wigner functions and fit each reconstructed distribution independently to the defect-state form in Eq.~\eqref{eqn:eqn2} to extract its centroid.
Figure~\ref{fig:figure3} shows the trajectories for $r_c=1$ and $2$ in both coupling regimes.
For both carrier amplitudes, the trajectories wind once around the origin for $g_1>g_2$ and have zero winding for $g_1<g_2$.
Changing $r_c$ rescales the trajectories while preserving these windings, as shown by Eq.~\eqref{eqn:eqn3}.
Therefore, the measurements resolve the phase-space invariant in each coupling regime while demonstrating control of its bosonic geometry.
The full distributions and fidelity analysis are given in Supplementary Sec.~\ref*{supp-sec:phase_space}
(Supplementary Figs.~\ref*{supp-fig:figure_motional_fidelity}--\ref*{supp-fig:figure_wigner_rc2}).

The Raman drives additionally generate a differential AC Stark term $\hat H_S/\hbar=\delta_S(t)\hat\sigma_z$~\cite{Haffner2003}.
This term breaks the chiral symmetry $\{\hat H,\hat\sigma_z\}=0$ of the ideal model and shifts the defect energy away from zero.
At fixed couplings, however, the ideal defect state is an exact eigenstate of $\hat\sigma_z$, so a static Stark term changes its energy without changing its wavefunction.
Its centroid and winding therefore remain unchanged within this Hamiltonian model.
During the preparation, the couplings and Stark shift vary in time; the resulting preparation errors are assessed in Appendix~\ref{Appendix_adiabatic_preparation}.

\begin{figure*}[t!]
\centering
\includegraphics[width=\linewidth]{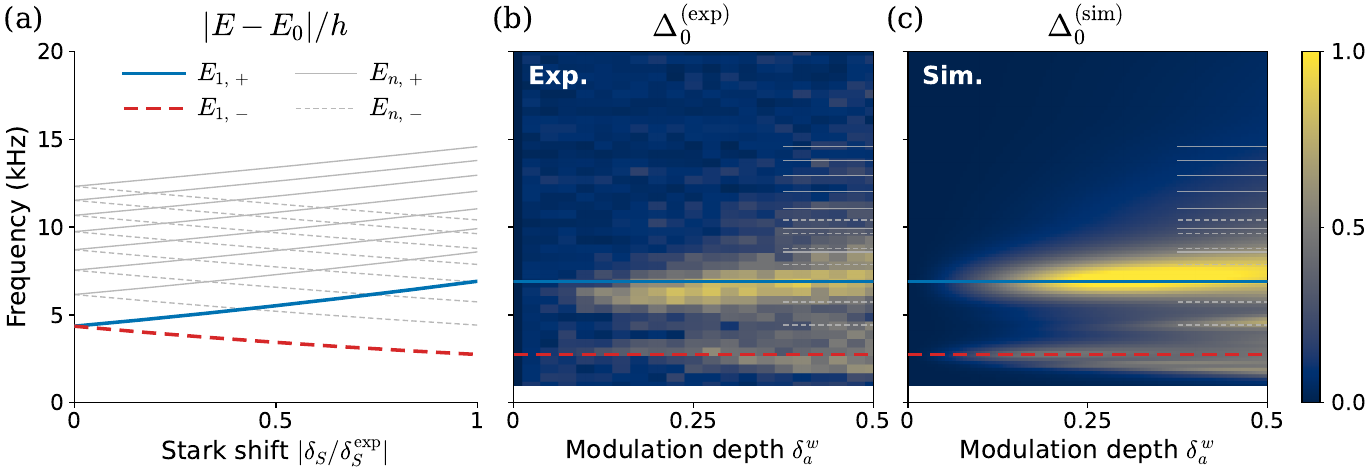}
\caption{
Carrier-modulation response of the defect state for $g_1>g_2$, $r_c=1$, $r_s=0.25$, and $\phi_{1,2}=0$. 
(a) Energy differences $|E_{n,\pm}-E_0|/h$ as the Stark shift increases to its experimental value. 
Blue solid and red dashed curves identify $n=1$, and gray curves show higher doublets. 
(b),(c) Measured and simulated RMS spin response $\Delta_0$ versus modulation frequency and modulation depth. 
The simulation includes the Stark shift and motional heating. Horizontal lines mark the $n=1$ gaps at the endpoint of (a). 
Gray ticks mark the energy differences $|E_{n,\pm}-E_0|/h$ for higher doublets ($n\ge2$), allowing their positions to be compared with the response features. 
These higher doublets are not directly coupled to the defect state by carrier modulation at first order.
}
\label{fig:figure4}
\end{figure*}

\subsection{Spin response to carrier modulation}
\label{sec:spin_robustness}

We probe the prepared defect state by applying carrier-amplitude modulation,
$w(t)=w[1+\delta_a^w\sin(2\pi f_Nt)]$, where $\delta_a^w$ is the fractional modulation depth and $f_N$ is the modulation frequency.
We set $w=g_{\max}$, $g_{\min}=g_{\max}/4$, and $\phi_{1,2}=0$.
The response is quantified by the RMS spin deviation from the unmodulated signal,
\begin{equation}
\Delta_0=\sqrt{\left\langle\left[\langle\hat\sigma_z(t)\rangle_{\mathrm{mod}}-\langle\hat\sigma_z(t)\rangle_0\right]^2\right\rangle_t},
\label{eqn:spin_response}
\end{equation}
averaged over $1\le f_Nt\le4$.
This observable quantifies modulation-induced changes in the
spin-sublattice populations.

To interpret the modulation response, we first consider the spectrum of
the unmodulated Hamiltonian $\hat H+\hbar\delta_S\hat\sigma_z$.
For fixed coupling parameters and Stark shift, the defect eigenstate $\ket{\psi_0}$ and excited eigenstates $\ket{\psi_{n,\pm}}$ have energies \(E_0=\hbar z_A\delta_S\) and \(E_{n,\pm}=\pm\hbar\sqrt{nG^2+\delta_S^2}\), respectively, where \(n\ge1\), $G=\sqrt{|g_1^2-g_2^2|}$, and $z_A=\langle A|\hat\sigma_z|A\rangle$.
The explicit forms of these eigenstates are given in Appendix~\ref{Appendix_adiabatic_preparation}.
Figure~\ref{fig:figure4}(a) shows the corresponding excitation frequencies
$|E_{n,\pm}-E_0|/h$ as $\delta_S$ varies from zero to its experimental value.

In the weak-modulation limit, carrier excitation is governed by $\langle\psi_{n,\pm}|\hat\sigma_x|\psi_0\rangle$.
Since $\hat U$ acts only on motion, the carrier flips $\ket A$ to $\ket B=\hat\sigma_x\ket A$ without changing the transformed oscillator state.
The resulting $\hat U\ket0\ket B$ belongs to the first excited doublet, so only $\ket{\psi_{1,+}}$ and $\ket{\psi_{1,-}}$ are directly coupled at first order.
Figure~\ref{fig:figure4} also shows the energy gaps from the defect state to higher doublets ($n\ge2$) for comparison with the measured response, although these doublets are not directly coupled to the defect state by carrier modulation at first order.
The two allowed channels have different matrix elements and spin contrasts when $\delta_S\ne0$, and therefore need not produce equally visible features in $\Delta_0$.

Figures~\ref{fig:figure4}(b) and (c) compare the measured response with a master-equation simulation including the Stark shift and motional heating.
The simulation reproduces the principal frequency-dependent features, including the unequal responses near the two first-doublet gaps.
At larger modulation depths, broadening and subsequent transitions beyond the directly coupled doublet can contribute, so the static gaps provide reference frequencies for the finite-drive response.
Additional parameter modulation data and transition-coupling calculations are given in Supplementary Sec.~\ref*{supp-sec:modulation} (Supplementary Figs.~\ref*{supp-fig:figure_defect_modulation}--\ref*{supp-fig:figure_response_stark_off}).
For the modulations and comparison states considered, the defect state has the fewest directly coupled final eigenstates at first order.
The finite energy gaps and restricted transition connectivity
help explain the stability of the defect-state spin polarization
away from the resonant frequencies~\cite{fsl_lee}.

\section{Discussion}

We have experimentally realized a topological defect state of the resonantly driven QRM using a single spin and a single motional mode of a trapped ion.
The centroid windings obtained from reconstructed motional phase-space distributions distinguish the two coupling regimes, linking the occupied spin sublattice of the nonuniform FSL to the geometry of the corresponding family of bosonic states.
Changing the carrier amplitude rescales the measured trajectories while preserving their winding, demonstrating control of the bosonic geometry within each coupling regime.

The response to modulation of the spin-drive amplitude complements this geometric characterization by probing the excitation structure of the defect state.
Finite energy gaps and restricted direct transition channels help explain the stability of its spin polarization away from resonance.
The static Stark term, proportional to $\sigma_z$, breaks chiral symmetry but leaves the ideal defect-state wavefunction unchanged since the state occupies a single spin sublattice.
For this perturbation, the centroid trajectory can therefore remain unchanged even as the excitation spectrum shifts.

Controlled light-shift compensation~\cite{Haffner2003,Stark_calib} would allow direct comparisons of the defect-state geometry and excitation response with and without the symmetry-breaking Stark effect.
Measurements approaching $g_1=g_2$ could further connect the reduction of the excitation gap to changes in the defect state's bosonic distribution.
Our results connect topological defect-state physics in a discrete Fock-state lattice to the oscillator's continuous phase-space geometry and to excitation dynamics probed through the spin.
Together, these results establish a single spin--oscillator pair as a minimal quantum system for realizing topological defect states and investigating their structure and stability.

\section*{Acknowledgments}
This work was supported by the National Research Foundation of Korea (NRF) Grant (No. RS-2024-00442855, No. RS-2024-00413957) and the Institute for Information \& Communications Technology Planning \& Evaluation (IITP) Grant (No. RS-2022-II221040), all funded by the Korean government (MSIT). 
S.Y. acknowledges support from the NRF Grant (No. RS-2025-19912971).

\section{Data availability}
The experimental and numerical data that support the findings of this paper are openly available in Zenodo~\cite{qrm_dataset}.

\section{Author contributions}
K.L. and J.K. contributed equally to this work. K.L. conceived the original idea and performed the numerical simulations. K.L. and J.K. performed the experiments, analysed the data and drafted the manuscript. J.Y. assisted with the experiments. S.Y. contributed to writing and revising the manuscript. W.C. contributed to discussions. T.K. supervised the project and secured funding.

\section{Competing interests}
The authors declare no competing interests.

\renewcommand{\bibsection}{\section*{References}}
\makeatletter
\def\@bibstyle{main}
\makeatother
\bibliography{main}

\clearpage

\appendix

\section*{Appendix}
\setcounter{secnumdepth}{1}

\section{Hamiltonian implementation and resonance conditions}
\label{Appendix_ion_trap_system}

The hyperfine qubit of a single $^{171}\mathrm{Yb}^{+}$ ion has splitting $\omega_{\mathrm{hf}}/(2\pi)=12.642812~\mathrm{GHz}$.
We use a radial mode at $\omega_x/(2\pi)=910~\mathrm{kHz}$ in a blade-shaped linear Paul trap~\cite{blade_setup,blade_setup2}; the other radial mode is separated by $360~\mathrm{kHz}$.
Sideband cooling gives $\bar n\simeq0.03$, with a measured heating rate of 5--6 quanta/s and motional coherence time of 2--4 ms.
Two perpendicular 355-nm Raman beams couple the spin and the selected motion.

The free laboratory-frame Hamiltonian is
\begin{equation}
\hat H_0/\hbar
=\omega_x\left(\hat a^\dagger\hat a+\frac12\right)
+\frac{\omega_{\mathrm{hf}}}{2}\hat\sigma_z.
\label{eqn:onsite_lab}
\end{equation}
The carrier, red-sideband, and blue-sideband Raman tones have frequencies
\begin{align}
\omega_c&=\omega_{\mathrm{hf}}+\delta_c,\nonumber\\
\omega_r&=\omega_{\mathrm{hf}}-\omega_x+\delta_r,\nonumber\\
\omega_b&=\omega_{\mathrm{hf}}+\omega_x+\delta_b,
\label{eqn:drive_detunings}
\end{align}
where $\delta_{c,r,b}$ denote their detunings from the respective resonances.

Transforming to $\ket{\psi_I}=e^{i\hat H_0t/\hbar}\ket{\psi_{\mathrm{lab}}}$ removes the free evolution.
The standard Lamb--Dicke expansion and rotating-wave approximation give the near-resonant interactions~\cite{RMP_iontrap}
\begin{align}
\hat H_c/\hbar
&=\frac{\Omega_c}{2}\hat\sigma_+e^{-i\delta_ct}
+\mathrm{H.c.},\nonumber\\
\hat H_r/\hbar
&=\frac{i\eta\Omega_r}{2}\hat a\hat\sigma_+
e^{-i\delta_rt+i\varphi_r}
+\mathrm{H.c.},\nonumber\\
\hat H_b/\hbar
&=\frac{i\eta\Omega_b}{2}\hat a^\dagger\hat\sigma_+
e^{-i\delta_bt+i\varphi_b}
+\mathrm{H.c.}
\label{eqn:sideband_transitions}
\end{align}
Here $\Omega_j$ and $\varphi_j$ are the two-photon Rabi frequency and Raman beat-note phase of tone $j$, respectively.
The carrier phase defines the spin reference, $\varphi_c=0$.
The Lamb--Dicke parameter is $\eta=k\sqrt{\hbar/(2m\omega_x)}$, where $k$ is the projection of the Raman wave-vector difference onto the selected motional axis.
The parameters in Eq.~\eqref{eqn:eqn1} are related to the Raman drives by
\begin{align}
w&=\frac{\Omega_c}{2},\qquad
g_1=\frac{\eta\Omega_r}{2},\qquad
g_2=\frac{\eta\Omega_b}{2},\nonumber\\
\phi_1&=-\varphi_r-\frac{\pi}{2},\qquad
\phi_2=-\varphi_b-\frac{\pi}{2}
\pmod{2\pi}.
\label{eqn:qrm_parameter_relation}
\end{align}
The experimental phase calibration includes these fixed offsets between the applied Raman phases and the interaction phases~\cite{ion_sdf,supplemental_material}.

For $\delta_c=(\delta_r+\delta_b)/2$, define
\begin{equation}
\omega_{\mathrm{eff}}
=\frac{\delta_r-\delta_b}{2},\qquad
\Omega_{\mathrm{eff}}
=-\frac{\delta_r+\delta_b}{2}.
\label{eqn:effective_onsite_coefficients}
\end{equation}
The transformation $\ket{\psi_{\mathrm{eff}}}=e^{-it\hat K}\ket{\psi_I}$, with $\hat K=\omega_{\mathrm{eff}}\hat a^\dagger\hat a +\Omega_{\mathrm{eff}}\hat\sigma_z/2$, makes all three interactions time independent.
Including the contribution $\hbar\hat K$ from this transformation gives the effective QRM Hamiltonian
\begin{align}
\hat H_{\mathrm{eff}}/\hbar
={}&\omega_{\mathrm{eff}}\hat a^\dagger\hat a
+\frac{\Omega_{\mathrm{eff}}}{2}\hat\sigma_z
+w\hat\sigma_x\nonumber\\
&+\left[
g_1e^{i\phi_1}\hat a^\dagger\hat\sigma_-
+g_2e^{i\phi_2}\hat a\hat\sigma_-
+\mathrm{H.c.}\right].
\label{eqn:qrm_with_onsite}
\end{align}
The effective FSL on-site energies are $\epsilon_{n,\uparrow/\downarrow}/\hbar = n\omega_{\mathrm{eff}}\pm\Omega_{\mathrm{eff}}/2$.
With all three tones resonant, $\delta_r=\delta_b=\delta_c=0$, both on-site coefficients vanish and Eq.~\eqref{eqn:qrm_with_onsite} reduces to Eq.~\eqref{eqn:eqn1}.
The physical trap frequency and qubit splitting remain finite.
The drive frequencies set the effective on-site energies in the rotating frame.
Off-resonant Raman couplings additionally generate $\hat H_S/\hbar=\delta_S\hat\sigma_z$, which is retained in the experimental model and analyzed in Supplementary Sec.~\ref*{supp-subsec:stark_shift} (Supplementary Fig.~\ref*{supp-fig:figure_stark_shift}).

\section{Defect-state form and adiabatic preparation}
\label{Appendix_adiabatic_preparation}

\begin{figure}[t!]
\centering
\includegraphics[width=\linewidth]
{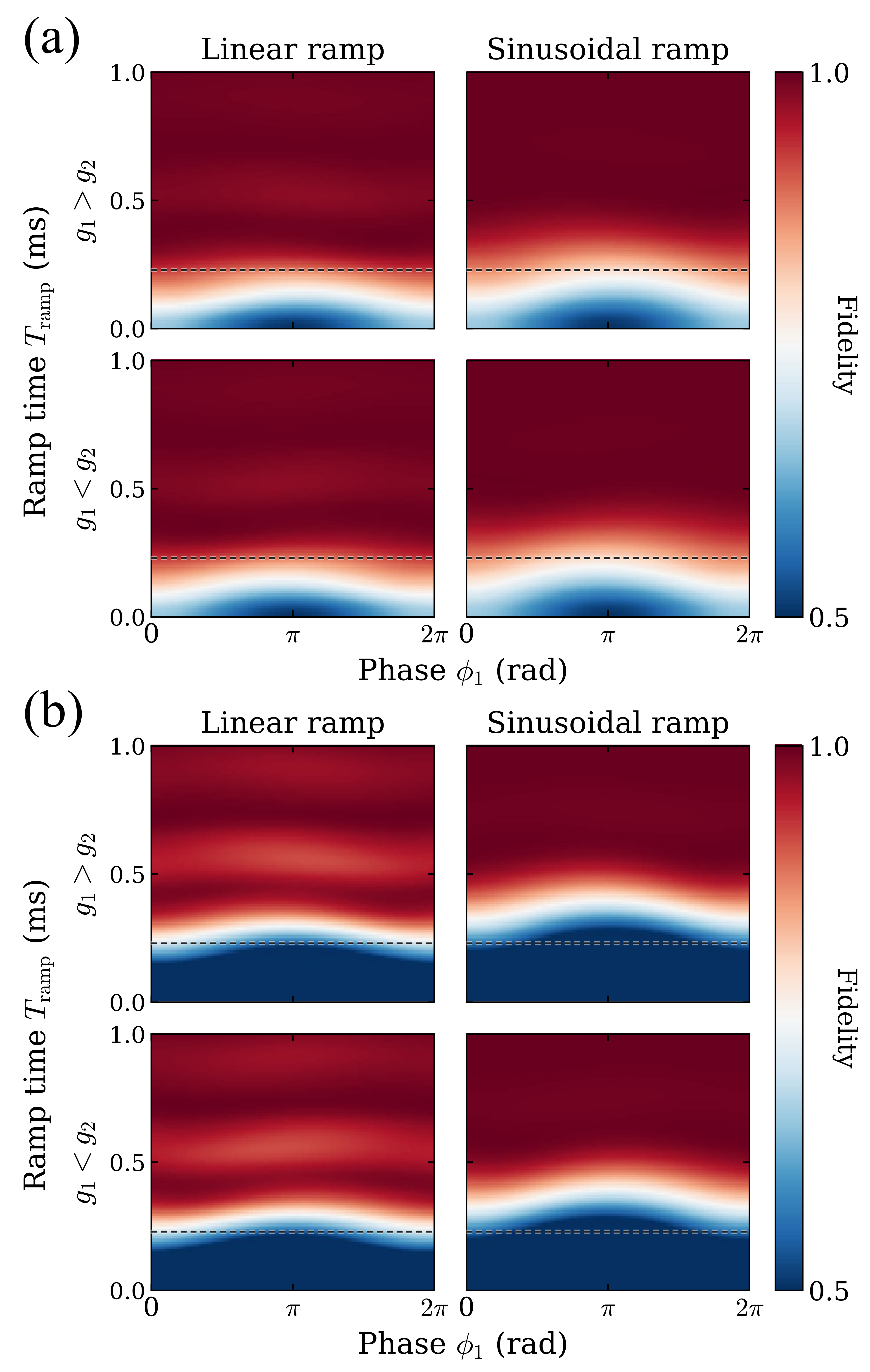}
\caption{Simulated defect-state preparation fidelity for
(a) $w_{\mathrm{target}}=g_{\max}$ and
(b) $w_{\mathrm{target}}=2g_{\max}$, versus ramp time and $\phi_1$,
at $\phi_2=0$ and
$g_{\min}^{\mathrm{target}}/g_{\max}=0.25$.
Linear and sinusoidal ramps follow Eq.~\eqref{eqn:eqnA2}.
Dashed lines mark the inverse ideal-gap scale
$229.5~\mu\mathrm{s}$.
The simulations include the drive-induced Stark shift with maximum
magnitude $|\delta_S|/(2\pi)\simeq2.23~\mathrm{kHz}$ during the ramp
and motional heating calibrated to a ground-state rate of
6 quanta/s.}
\label{fig:figure_adiabatic}
\end{figure}

The operators in Eq.~\eqref{eqn:eqn2} are $\hat R=e^{-i\theta\hat a^\dagger\hat a}$, $\hat D=e^{\alpha\hat a^\dagger-\alpha^*\hat a}$, and $\hat S=e^{r(\hat a^2-\hat a^{\dagger2})/2}$~\cite{quantum_optics,phase_shift_operator}.
Their parameters are~\cite{fsl_lee}
\begin{align}
\alpha
&=\frac{w(g_2e^{-i\phi}-g_1e^{i\phi})}
{g_1^2-g_2^2},\qquad
\tanh(2r)=\frac{2g_1g_2}{g_1^2+g_2^2},
\nonumber\\
2\phi&=\phi_1+\phi_2,\qquad
2\theta=\phi_2-\phi_1.
\label{eqn:defect_parameters}
\end{align}
The occupied sublattice is identified by $\ket A=\ket\down$ for $g_1>g_2$ and $\ket A=\ket\up$ for $g_1<g_2$.
We denote the opposite spin state by $\ket B$ and choose its relative phase so that the coupling between $\hat U\ket n\ket A$ and $\hat U\ket{n-1}\ket B$ is real and positive.
For $\phi_1=\phi_2=0$, as in the carrier-modulation measurement, this convention gives $\ket B=\hat\sigma_x\ket A$.

The unpaired defect state is $\ket{\psi_0}=\hat U\ket0\ket A$.
For $n\ge1$, the normalized eigenstates of the ideal Hamiltonian are~\cite{fsl_lee}
\begin{align}
\ket{\psi_{n,\pm}^{(0)}}
&=\frac{\hat U}{\sqrt2}
\left(\ket n\ket A
\pm\ket{n-1}\ket B\right),\nonumber\\
E_{n,\pm}^{(0)}
&=\pm\hbar G\sqrt n,\qquad
G=\sqrt{|g_1^2-g_2^2|}.
\label{eqn:eqnA1}
\end{align}
The gap from the defect state to the first doublet is $\Delta E=\hbar G$.

For a fixed differential Stark shift $\hat H_S=\hbar\delta_S\hat\sigma_z$, the unpaired state retains the same wavefunction and acquires energy $E_0=\hbar z_A\delta_S$, where $z_A=\langle A|\hat\sigma_z|A\rangle$.
Within each $n\ge1$ doublet, the normalized eigenstates become
\begin{align}
\ket{\psi_{n,+}}
&=\hat U\left(
c_n\ket n\ket A
+s_n\ket{n-1}\ket B\right),\nonumber\\
\ket{\psi_{n,-}}
&=\hat U\left(
s_n\ket n\ket A
-c_n\ket{n-1}\ket B\right),
\label{eqn:stark_eigenstates}
\end{align}
where
\begin{align}
\lambda_n&=\sqrt{nG^2+\delta_S^2},\nonumber\\
c_n&=\sqrt{\frac{\lambda_n+z_A\delta_S}
{2\lambda_n}},\qquad
s_n=\sqrt{\frac{\lambda_n-z_A\delta_S}
{2\lambda_n}}.
\label{eqn:stark_mixing}
\end{align}
Their energies are $E_{n,\pm}=\pm\hbar\lambda_n$, as stated in the main text.
At $\delta_S=0$, Eq.~\eqref{eqn:stark_eigenstates} reduces to the ideal eigenstates in Eq.~\eqref{eqn:eqnA1}.
As $\hat U$ acts only on motion, a carrier spin flip from $\ket{\psi_0}$ lies in the $n=1$ doublet. For weak carrier modulation there is no direct coupling from the defect state to higher doublets.

For preparation, we switch on the stronger sideband at $g_{\max}/(2\pi)=4.5~\mathrm{kHz}$ and ramp the remaining couplings
from zero to $g_{\min}^{\mathrm{target}}=0.25g_{\max}$ and $w_{\mathrm{target}}=g_{\max}$ or $2g_{\max}$.
We compare linear and sinusoidal profiles,
\begin{align}
w(t)&=w_{\mathrm{target}}s(t),\qquad
g_{\min}(t)=g_{\min}^{\mathrm{target}}s(t),
\nonumber\\
s(t)&=\begin{cases}
t/T_{\mathrm{ramp}},&\text{linear},\\
\sin^2[\pi t/(2T_{\mathrm{ramp}})],&\text{sinusoidal}.
\end{cases}
\label{eqn:eqnA2}
\end{align}
At the endpoint, the inverse ideal-gap scale is $2\pi/G\simeq229.5~\mu\mathrm{s}$.
This provides a reference timescale rather than a sufficient adiabaticity criterion: excitation also depends on the ramp matrix elements and the instantaneous gaps, including the Stark shift.

Master-equation simulations using QuTiP~\cite{qutip} include the time-dependent Stark shift generated during the ramp and motional heating through a term proportional to $\mathcal D[\hat a^\dagger]\rho$, calibrated to a ground-state heating rate of 6 quanta/s.
Figure~\ref{fig:figure_adiabatic} evaluates the fidelity with the ideal target state in Eq.~\eqref{eqn:eqn2}.
The sinusoidal profile has vanishing slopes at both endpoints and generally yields higher preparation fidelity than the linear profile in the simulated parameter range.
We use it experimentally with $T_{\mathrm{ramp}}=0.5~\mathrm{ms}$ for Figs.~\ref{fig:figure2} and \ref{fig:figure3}, and $T_{\mathrm{ramp}}=1~\mathrm{ms}$ for Fig.~\ref{fig:figure4}.

\end{document}


\title{
Supplementary Information: Observation of a topological defect state in the quantum Rabi model
}

\author{Kyungmin Lee}
\thanks{These authors contributed equally.}
\affiliation{\mbox{Department of Computer Science and Engineering, Seoul National University, Seoul 08826, Republic of Korea}}
\affiliation{\mbox{Automation and System Research Institute, Seoul National University, Seoul 08826, Republic of Korea}}
\affiliation{\mbox{NextQuantum, Seoul National University, Seoul 08826, Republic of Korea}}

\author{Jiyong Kang}
\thanks{These authors contributed equally.}
\affiliation{\mbox{Department of Computer Science and Engineering, Seoul National University, Seoul 08826, Republic of Korea}}
\affiliation{\mbox{Automation and System Research Institute, Seoul National University, Seoul 08826, Republic of Korea}}
\affiliation{\mbox{NextQuantum, Seoul National University, Seoul 08826, Republic of Korea}}

\author{Sunkyu Yu}
\affiliation{Intelligent Wave Systems Laboratory, Department of Electrical and Computer Engineering, Seoul National University, Seoul 08826, Republic of Korea}

\author{Jaehun You}
\affiliation{\mbox{Department of Computer Science and Engineering, Seoul National University, Seoul 08826, Republic of Korea}}
\affiliation{\mbox{Automation and System Research Institute, Seoul National University, Seoul 08826, Republic of Korea}}
\affiliation{\mbox{NextQuantum, Seoul National University, Seoul 08826, Republic of Korea}}

\author{Wonhyeong Choi}
\affiliation{\mbox{Department of Computer Science and Engineering, Seoul National University, Seoul 08826, Republic of Korea}}
\affiliation{\mbox{Automation and System Research Institute, Seoul National University, Seoul 08826, Republic of Korea}}
\affiliation{\mbox{NextQuantum, Seoul National University, Seoul 08826, Republic of Korea}}

\author{Taehyun Kim}
\email{Corresponding author: taehyun@snu.ac.kr}
\affiliation{\mbox{Department of Computer Science and Engineering, Seoul National University, Seoul 08826, Republic of Korea}}
\affiliation{\mbox{Automation and System Research Institute, Seoul National University, Seoul 08826, Republic of Korea}}
\affiliation{\mbox{NextQuantum, Seoul National University, Seoul 08826, Republic of Korea}}
\affiliation{\mbox{Institute of Applied Physics, Seoul National University, Seoul 08826, Republic of Korea}}
\affiliation{\mbox{Inter-university Semiconductor Research Center, Seoul National University, Seoul 08826, Republic of Korea}}
\affiliation{\mbox{Institute of Computer Technology, Seoul National University, Seoul 08826, Republic of Korea}}

\maketitle
\tableofcontents

\section{Calibration of the sideband phases}
\label{sec:phase_calibration}

\begin{figure*}[b!]
\centering
\includegraphics[width=\textwidth]{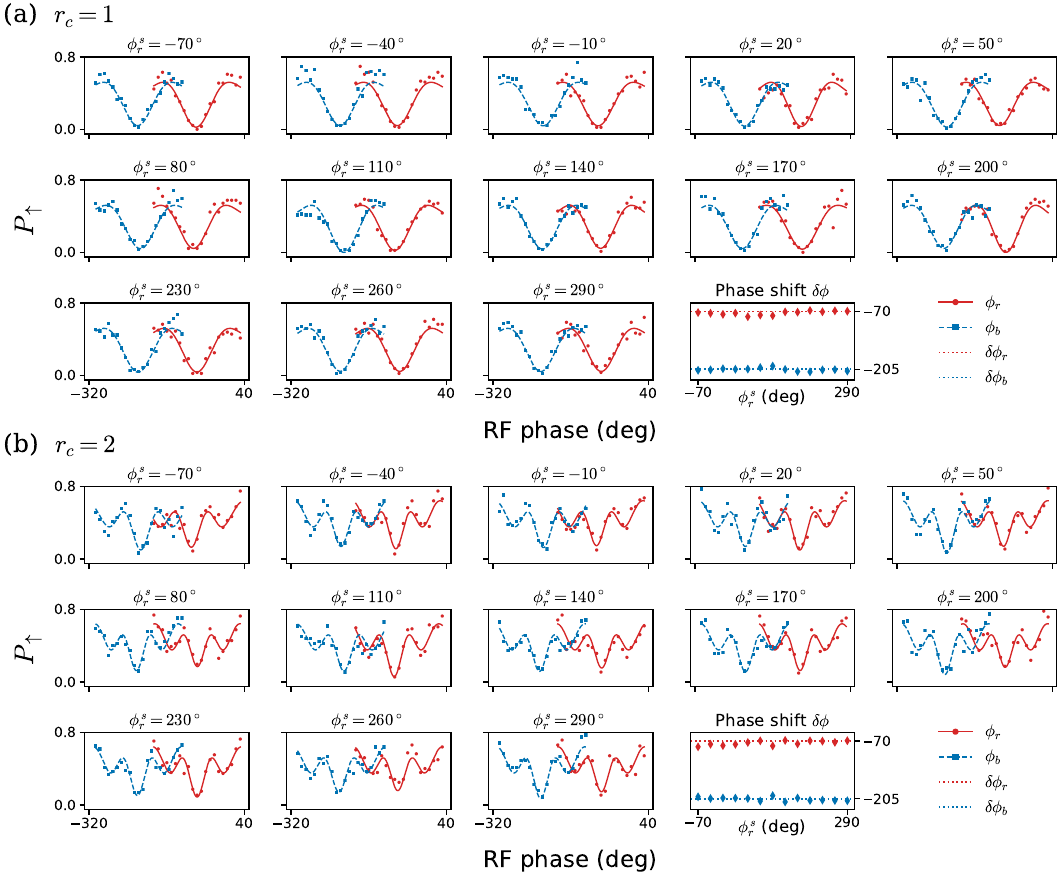}
\caption{
Sideband-phase calibration for (a) \(r_c=1\) and (b) \(r_c=2\).
Each small panel corresponds to the indicated target red-sideband RF phase \(\phi_r^s\) and shows \(P_{\up}\) while scanning the red-sideband phase (red circles) or blue-sideband phase (blue squares), with the carrier phase fixed.
Curves are fits to Eq.~\eqref{eqn:phase_calibration}.
The phase-offset panel in each block shows the fitted phase offsets versus \(\phi_r^s\); horizontal dotted lines mark the offsets used in the experiment, \(\delta\phi_r=-70^\circ\) and \(\delta\phi_b=-205^\circ\).
Error bars in the phase-offset panels denote one-standard-deviation uncertainties in the fitted phase offsets.
}
\label{fig:figure_calibration}
\end{figure*}

The carrier amplitude and the two sideband amplitudes are set by the corresponding radio-frequency (RF) amplitudes, whereas their relative optical phases contain fixed offsets from the RF and optical paths.
We determine these offsets before the characteristic-function measurements by preparing a defect state in a single-sideband limit and canceling its coherent displacement with a calibrated spin-dependent displacement~\cite{ion_sdf}.
The Hamiltonian phases are related to the applied Raman phases by Eq.~\textup{(\ref*{main-eqn:qrm_parameter_relation})} of the main text.

With only the red sideband present, the defect state occupies \(\ket{\down}\) and reduces to
\begin{equation}
\ket{\psi_0^{(r)}}
=\ket{\gamma_r}\ket{\down},
\qquad
\gamma_r=-\frac{w}{g_1}e^{i\phi_1}.
\label{eqn:red_calibration_state}
\end{equation}
With only the blue sideband present, it occupies \(\ket{\up}\) and becomes
\begin{equation}
\ket{\psi_0^{(b)}}
=\ket{\gamma_b}\ket{\up},
\qquad
\gamma_b=-\frac{w}{g_2}e^{-i\phi_2}.
\label{eqn:blue_calibration_state}
\end{equation}

We perform the calibration at both \(r_c=1\) and \(r_c=2\). 
For the red-sideband calibration, we use \(g_1= (2\pi)~4.5~\mathrm{kHz}\) and \(w=r_cg_1\). 
For the blue-sideband calibration, we use \(g_2=(2\pi)~4.5~\mathrm{kHz}\) and \(w=r_cg_2\), and a microwave \(\pi\) pulse maps the spin to the same readout convention used for the red-sideband case.

After the state preparation, we apply a calibrated spin-dependent displacement~\cite{ion_sdf}.
At the phase that gives perfect cancellation, the motional state returns to the vicinity of the vacuum and a subsequent red-sideband probe gives a minimum spin-transfer probability.
We write the residual displacement during an RF-phase scan as
\begin{equation}
\beta(\phi)=\alpha_{\mathrm{cal}}
\left[1-(1-\epsilon)e^{i(\phi-\phi_{\mathrm{shift}})}\right],
\label{eqn:residual_displacement}
\end{equation}
In the fit, the effective magnitude \(|\alpha_{\mathrm{cal}}|\), the offset \(\phi_{\mathrm{shift}}\), and \(\epsilon\), which allows a small mismatch between the state-preparation and analysis displacements, are free parameters.
The corresponding coherent-state phonon distribution is
\begin{equation}
P_n(\phi)=e^{-|\beta(\phi)|^2}
\frac{|\beta(\phi)|^{2n}}{n!}.
\label{eqn:residual_phonon_population}
\end{equation}
For a red-sideband probe whose duration gives a \(\pi\) pulse on the \(n=1\) transition, the measured population is~\cite{RMP_iontrap}
\begin{equation}
P_{\up}(\phi)=
\sum_{n=1}^{n_{\max}}P_n(\phi)
\sin^2\!\left(\frac{\pi}{2}\sqrt{n}\right),
\qquad n_{\max}=30.
\label{eqn:phase_calibration}
\end{equation}

Figure~\ref{fig:figure_calibration} shows the calibration scans acquired for the two trajectory radii used in the phase-space measurement.
The extracted offsets depend only weakly on the target red-sideband phase.
We therefore apply the fixed offsets \(\delta\phi_r=-70^\circ\) and \(\delta\phi_b=-205^\circ\) during state preparation for all trajectory measurements.
These empirical offsets include the fixed phase conversion in Eq.~\textup{(\ref*{main-eqn:qrm_parameter_relation})} and define the nominal Hamiltonian phases \(\phi_1\) and \(\phi_2\) used in the main text.

\section{Stark shift and response to parameter modulation}
\label{sec:modulation}

\begin{figure}[b!]
\centering
\includegraphics[width=0.75\linewidth]{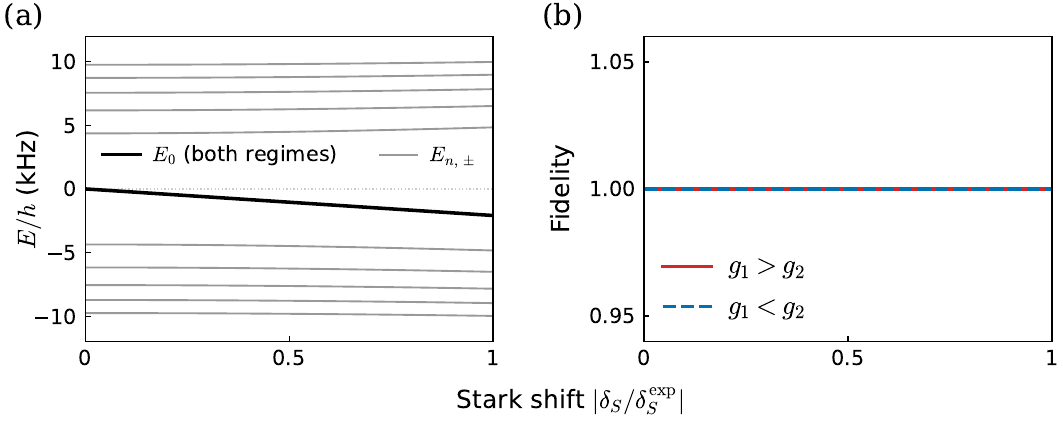}
\caption{
Static effect of the differential AC Stark shift.
(a) Eigenenergies expressed as \(E/h\) (in kilohertz) versus the Stark-shift magnitude normalized by its experimental endpoint.
The black curve is the defect level \(E_0\), which is the same for the two coupling regimes after their correlated reversals of \(z_A\) and \(\delta_S\).
Gray curves show higher doublets \(E_{n,\pm}\).
(b) Fidelity of the Stark-shifted defect eigenstate with the ideal defect state for \(g_1>g_2\) (red) and \(g_1<g_2\) (blue).
The coincident unit-fidelity curves show that the fixed Stark term leaves the exact defect-state wavefunction unchanged.
}
\label{fig:figure_stark_shift}
\end{figure}

\subsection{Differential AC Stark shift}
\label{subsec:stark_shift}

The red- and blue-sideband Raman tones are detuned from the carrier transition by approximately \(-\omega_x\) and \(+\omega_x\), respectively.
Their off-resonant carrier couplings generate the differential Stark term
\begin{equation}
\hat H_S/\hbar=\delta_S\hat\sigma_z,
\qquad
\delta_S\simeq\frac{\Omega_r^2-\Omega_b^2}{4\omega_x}
=\frac{g_1^2-g_2^2}{\eta^2\omega_x},
\label{eqn:stark_shift}
\end{equation}
where \(g_{1,2}=\eta\Omega_{r,b}/2\)~\cite{RMP_iontrap,Haffner2003}.
For \(\eta=0.1\), \(\omega_x=(2\pi)~910~\mathrm{kHz}\), and \(g_{\max}=(2\pi)~4.5~\mathrm{kHz}\), the limiting value at \(g_{\min}=0\) is \(|\delta_S|=(2\pi)~2.23~\mathrm{kHz}\).
At the experimental ratio \(g_{\min}/g_{\max}=0.25\), it is
\begin{equation}
\frac{|\delta_S^{\mathrm{exp}}|}{2\pi}=2.09~\mathrm{kHz}.
\label{eqn:experimental_stark_shift}
\end{equation}

The sign of \(\delta_S\) reverses when the stronger sideband is exchanged.
At the same time, the occupied spin sublattice changes from \(\ket{\down}\) for \(g_1>g_2\) to \(\ket{\up}\) for \(g_1<g_2\).
Consequently, the defect-state energy
\begin{equation}
E_0=\hbar z_A\delta_S,
\qquad z_A=\bra A\hat\sigma_z\ket A,
\label{eqn:stark_defect_energy}
\end{equation}
decreases in both regimes for the experimentally generated Stark shift, as shown in Fig.~\ref{fig:figure_stark_shift}(a).
The excited-state energies are
\begin{equation}
E_{n,\pm}=\pm\hbar\sqrt{nG^2+\delta_S^2},
\qquad
G=\sqrt{|g_1^2-g_2^2|}.
\label{eqn:stark_excited_energies}
\end{equation}
Since the ideal defect state is an eigenstate of \(\hat\sigma_z\), adding a static \(\delta_S\hat\sigma_z\) term changes its energy but not its wavefunction.
Thus the two fidelity curves in Fig.~\ref{fig:figure_stark_shift}(b) overlap at unity.
This static effect does not remove preparation errors produced while the couplings and the Stark shift vary during the ramp; those errors are evaluated in the Appendix of the main text.

\subsection{Modulation protocols and numerical model}
\label{subsec:modulation_protocols}

For the modulation measurements and simulations we use
\begin{equation}
w=g_{\max}=(2\pi)~4.5~\mathrm{kHz},
\qquad
g_{\min}=0.25g_{\max},
\qquad
\phi_1=\phi_2=0.
\label{eqn:modulation_parameters}
\end{equation}
We apply three sinusoidal perturbations at modulation frequency \(f_N\).
In weaker-sideband frequency modulation, the instantaneous sideband-frequency offset has amplitude \(\delta_f^g\), so the accumulated interaction phase is
\begin{align}
\phi_{\min}(t)
&=\phi_{\min}(0)
+2\pi\int_0^t\!\delta_f^g\sin(2\pi f_Nt')\,dt'\nonumber\\
&=\phi_{\min}(0)
+\frac{\delta_f^g}{f_N}\left[1-\cos(2\pi f_Nt)\right].
\label{eqn:frequency_modulation}
\end{align}
The other two protocols are weaker-sideband amplitude modulation,
\begin{equation}
g_{\min}(t)=g_{\min}
\left[1+\delta_a^g\sin(2\pi f_Nt)\right],
\label{eqn:sideband_amplitude_modulation}
\end{equation}
and carrier-amplitude modulation,
\begin{equation}
w(t)=w\left[1+\delta_a^w\sin(2\pi f_Nt)\right].
\label{eqn:carrier_amplitude_modulation}
\end{equation}
In Eq.~\eqref{eqn:sideband_amplitude_modulation}, the instantaneous Stark shift is recalculated from Eq.~\eqref{eqn:stark_shift}, while it is constant for the other two protocols.

The simulations solve
\begin{equation}
\dot\rho=-\frac{i}{\hbar}
[\hat H(t)+\hat H_S(t),\rho]
+\mathcal D[\hat L_h]\rho,
\qquad
\hat L_h=\sqrt{\Gamma_h}\,\hat a^\dagger,
\label{eqn:modulation_master_equation}
\end{equation}
using QuTiP~\cite{qutip}, a motional cutoff of 30, and \(\Gamma_h=6~\mathrm{s}^{-1}\), which corresponds to the heating rate at the ground state.
The chosen jump operator gives
\(d\langle n\rangle/dt=\Gamma_h(\langle n\rangle+1)\), rather than a state-independent rate of six quanta per second.
For every frequency and modulation depth, the response is evaluated over the same dimensionless time window \(1\le f_Nt\le4\) and compared with an unmodulated trace evolved for the same physical duration.
For the defect-state response simulations, the modulation evolution starts from the ideal defect state at the target couplings. 
Preparation errors during the ramp are assessed separately in the main-text Appendix.

\begin{figure*}[t!]
\centering
\includegraphics[width=\textwidth]{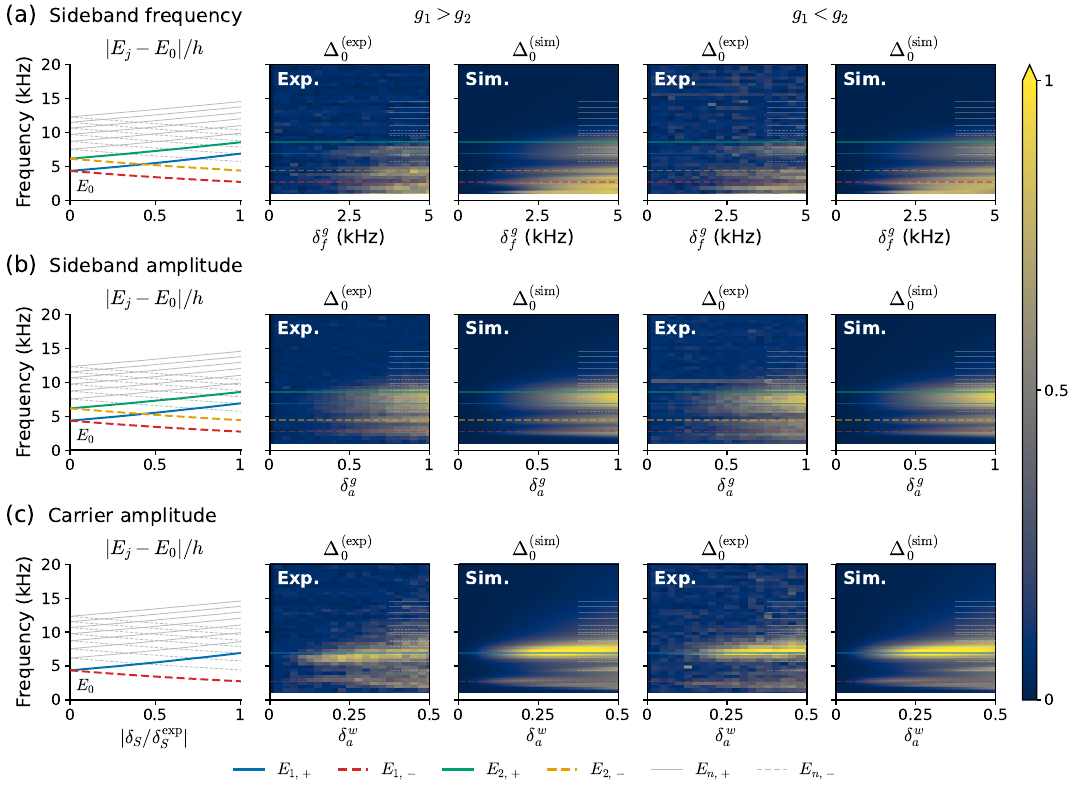}
\caption{
Defect-state spin response for the experimental Stark shift.
Rows show (a) weaker-sideband frequency modulation, (b) weaker-sideband amplitude modulation, and (c) carrier-amplitude modulation.
The first panel in each row shows \(|E_j-E_0|/h\) as the Stark shift increases from zero to \(|\delta_S^{\mathrm{exp}}|\).
The remaining panels show the measured and simulated \(\Delta_0\) for \(g_1>g_2\) and \(g_1<g_2\).
Colored horizontal lines identify low-lying directly coupled levels.
Short gray lines mark higher gaps as spectral references.
The common linear color scale runs from 0 to 1.
}
\label{fig:figure_defect_modulation}
\end{figure*}

\begin{figure*}[t!]
\centering
\includegraphics[width=\textwidth]{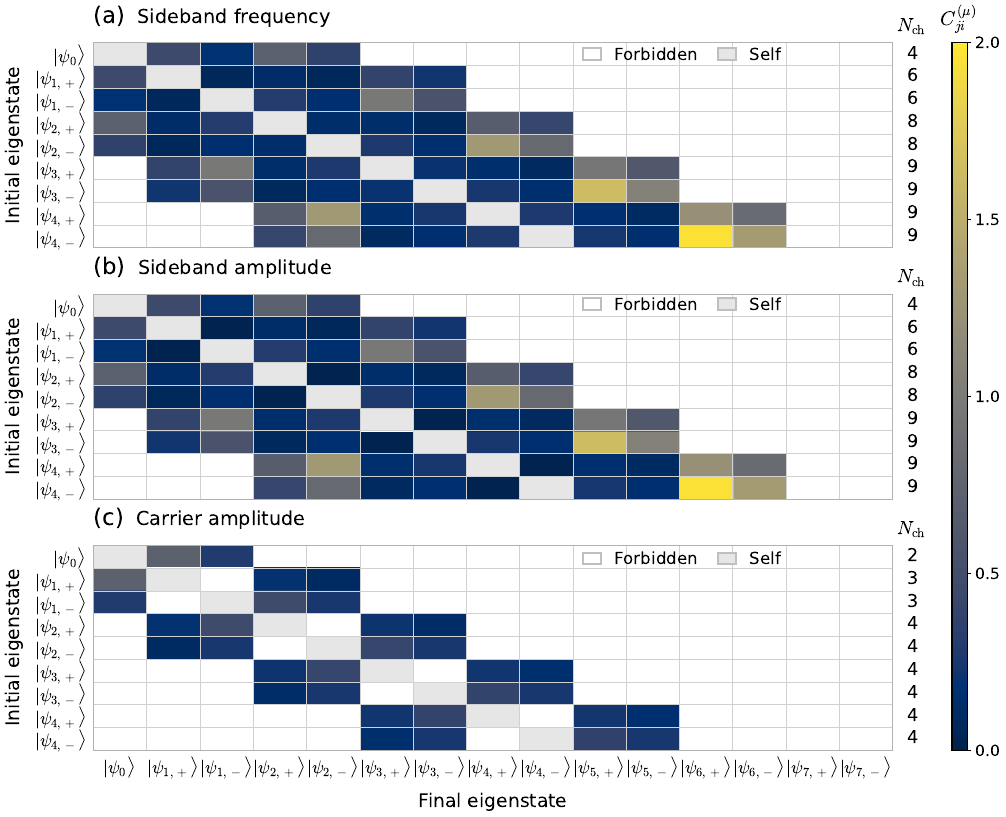}
\caption{
First-order transition-coupling matrices in the presence of the experimental Stark shift.
Rows of panels correspond to (a) weaker-sideband phase (the coupling underlying sideband-frequency modulation), (b) weaker-sideband amplitude, and (c) carrier amplitude.
Rows within each matrix are initial eigenstates through \(n=4\), and columns are final eigenstates through \(n=7\).
Colored cells show \(C_{ji}^{(\mu)}\) from Eq.~\eqref{eqn:transition_coupling}.
White cells are forbidden at first order and gray cells are self matrix elements, which are excluded from \(N_{\mathrm{ch}}\).
The matrices are identical for the two coupling regimes after exchanging the spin sublattices and are therefore shown once.
All panels use the same fixed linear scale, 0--2.
}
\label{fig:figure_transition_stark_on}
\end{figure*}

\begin{figure*}[t!]
\centering
\includegraphics[width=\textwidth]{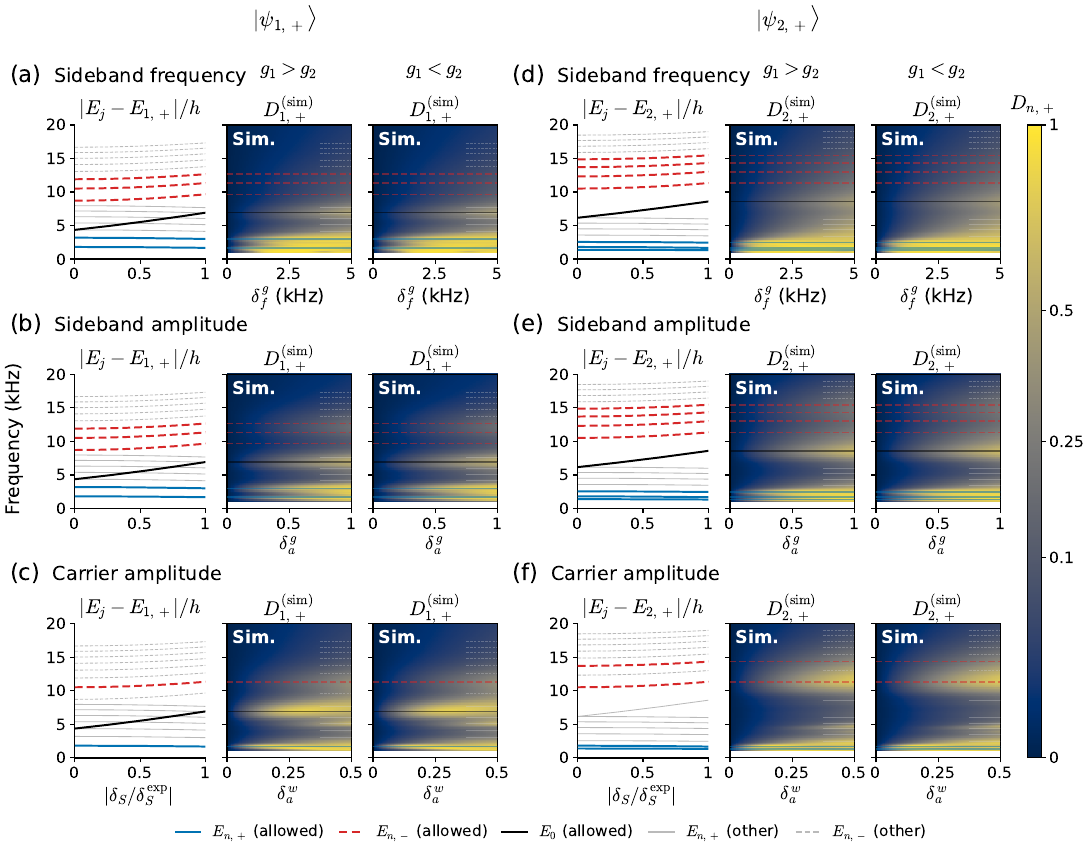}
\caption{
Survival response of the finite-energy eigenstates \(\ket{\psi_{1,+}}\) (left block) and \(\ket{\psi_{2,+}}\) (right block) in the presence of the experimental Stark shift.
Rows show weaker-sideband frequency, weaker-sideband amplitude, and carrier-amplitude modulation.
Within each block, the first panel shows the static gaps from the initial state, and the next two panels show \(D_{n,+}\) for \(g_1>g_2\) and \(g_1<g_2\).
Colored curves and horizontal lines mark first-order allowed transitions; gray curves and short gray lines show other spectral gaps.
Every response map uses the same 0--1 color range with a power-law normalization of exponent \(1/2\).
}
\label{fig:figure_eigenstate_response}
\end{figure*}

\begin{figure*}[t!]
\centering
\includegraphics[width=\textwidth]{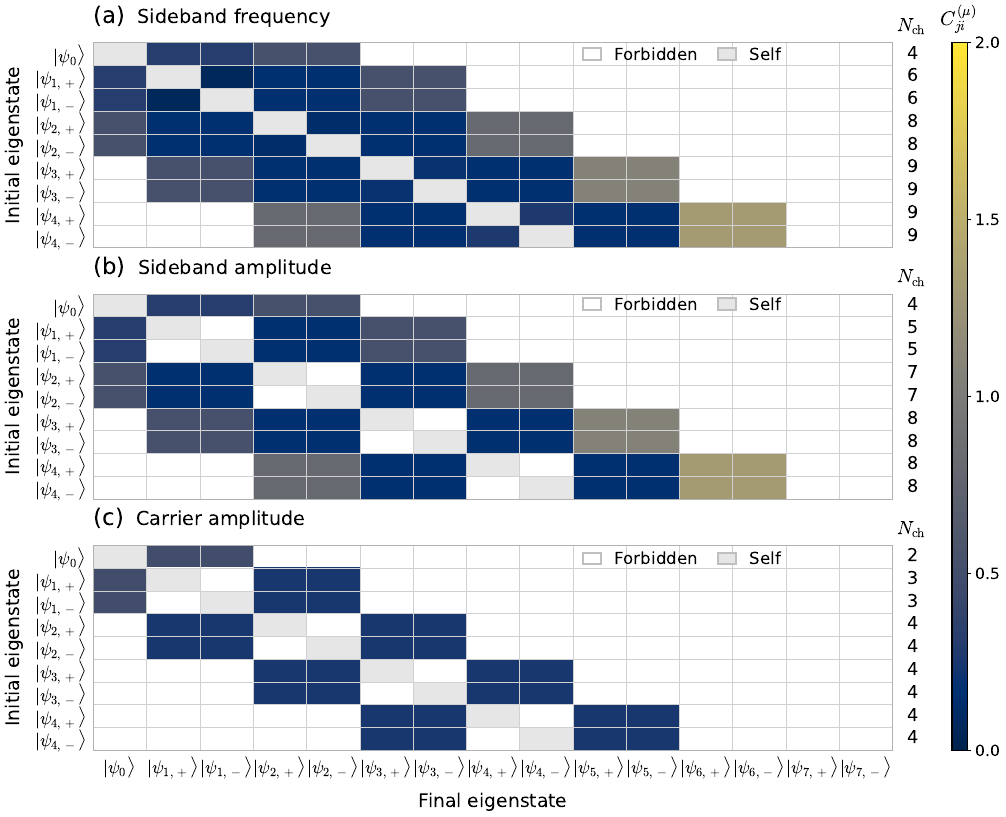}
\caption{
First-order transition-coupling matrices with the differential Stark term removed.
The panel, state, and \(N_{\mathrm{ch}}\) conventions are the same as in Fig.~\ref{fig:figure_transition_stark_on}.
The color normalization is also shared with that figure, so the Stark-on and Stark-off coupling strengths can be compared directly.
In panel (b), changes in the eigenstate mixing and the removal of the Stark-shift derivative from the perturbation operator modify both the coupling strengths and the number of allowed channels.
}
\label{fig:figure_transition_stark_off}
\end{figure*}

\begin{figure*}[t!]
\centering
\includegraphics[width=\textwidth]{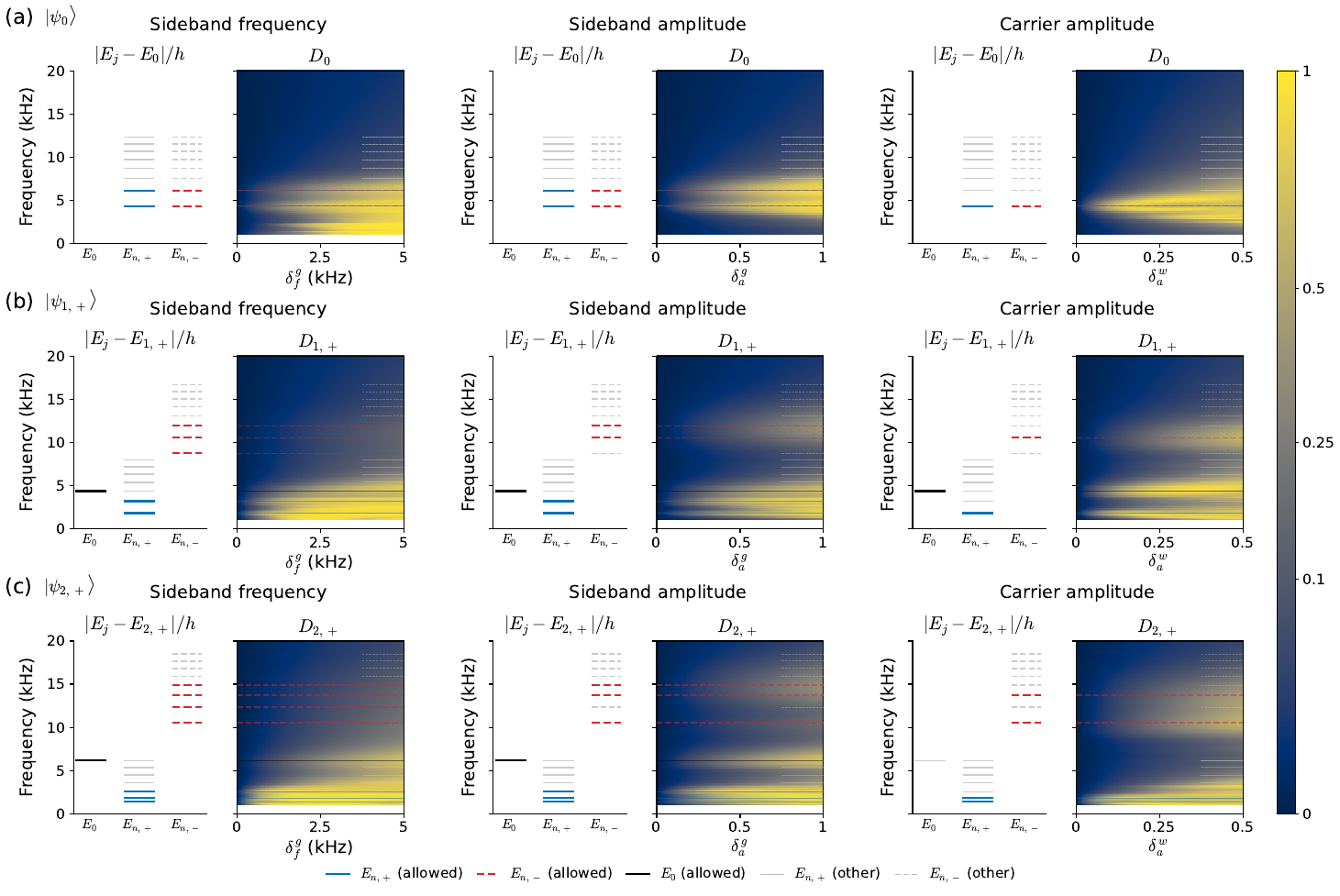}
\caption{
Stark-shift-free survival response for (a) \(\ket{\psi_0}\), (b) \(\ket{\psi_{1,+}}\), and (c) \(\ket{\psi_{2,+}}\) in the \(g_1>g_2\) regime.
Columns correspond to weaker-sideband frequency, weaker-sideband amplitude, and carrier-amplitude modulation.
For each protocol, the left panel shows the fixed excitation spectrum \(|E_j-E_i|/h\) and the right panel shows \(D_i\).
Colored levels and horizontal guides identify first-order allowed transitions, and gray levels denote the remaining gaps.
The response maps use the same 0--1 power-law color normalization as Fig.~\ref{fig:figure_eigenstate_response}.
}
\label{fig:figure_response_stark_off}
\end{figure*}

\subsection{Defect-state spin response}
\label{subsec:defect_response}

The experimental observable is the root-mean-square change of the spin polarization,
\begin{equation}
\Delta_0=
\sqrt{\left\langle
\left[
\langle\hat\sigma_z(t)\rangle_{\mathrm{mod}}
-\langle\hat\sigma_z(t)\rangle_0
\right]^2
\right\rangle_{1\le f_Nt\le4}},
\label{eqn:defect_spin_response}
\end{equation}
where \(\langle\hat\sigma_z(t)\rangle_0\) denotes evolution with zero modulation depth.
Thus \(\Delta_0\) measures a change in the spin-sublattice populations. 

Figure~\ref{fig:figure_defect_modulation} compares the measured and simulated \(\Delta_0\) for all three protocols and for both coupling regimes.
The simulation reproduces the principal frequency-dependent structures of the measured response.
The spectrum at the left of each row shows the excitation frequencies as the Stark shift increases to its experimental value, while the horizontal lines in the response maps mark the endpoint frequencies.
These lines are guides for interpreting the weak-drive response, not predictions that every line must be resolved.
At finite modulation depth, matrix-element variations, power broadening, the frequency-dependent phase excursion in Eq.~\eqref{eqn:frequency_modulation}, heating, and multistep transitions all affect the observed contrast.

\subsection{First-order transition couplings}
\label{subsec:transition_couplings}

To separate static connectivity from the finite-drive response, we calculate the perturbation matrix elements in the eigenbasis of \(\hat H+\hat H_S\).
Writing \(\hat K=(\hat H+\hat H_S)/\hbar\), an infinitesimal perturbation of type \(\mu\) is parameterized as
\begin{equation}
\delta\hat K_\mu=g_{\mathrm{ref}}\epsilon_\mu\hat O_\mu,
\qquad
g_{\mathrm{ref}}=(2\pi)~4.5~\mathrm{kHz}.
\label{eqn:coupling_normalization}
\end{equation}
We use one fixed normalization for every initial state, modulation type, coupling regime, and Stark condition:
\begin{align}
\epsilon_w&=\frac{\delta w}{g_{\mathrm{ref}}},
&\hat O_w&=\frac{\partial\hat K}{\partial w}=\hat\sigma_x,
\nonumber\\
\epsilon_g&=\frac{\delta g_{\min}}{g_{\mathrm{ref}}},
&\hat O_g&=\frac{\partial\hat K}{\partial g_{\min}},
\nonumber\\
\epsilon_\phi&=\frac{g_{\min}\delta\phi_{\min}}{g_{\mathrm{ref}}},
&\hat O_\phi&=\frac{1}{g_{\min}}
\frac{\partial\hat K}{\partial\phi_{\min}}.
\label{eqn:modulation_operators}
\end{align}
For \(\hat O_g\), the derivative includes the amplitude dependence of \(\delta_S\) in Eq.~\eqref{eqn:stark_shift}.
For sideband-frequency modulation, \(\hat O_\phi\) is the local phase-coupling operator, while the actual time-dependent phase amplitude is set by Eq.~\eqref{eqn:frequency_modulation}.

The plotted dimensionless coupling strength is
\begin{equation}
C_{ji}^{(\mu)}=
\left|\braket{\psi_j|\hat O_\mu|\psi_i}\right|^2.
\label{eqn:transition_coupling}
\end{equation}
No row-wise or panel-wise normalization is applied.
The number \(N_{\mathrm{ch}}\) counts distinct final eigenstates with a nonzero off-diagonal matrix element from the indicated initial state.
It excludes the initial state itself, counts degenerate final eigenstates separately, and is not restricted to the 0--20-kHz response window.
The final-state range through \(n=7\) contains all first-order channels from the displayed initial states through \(n=4\).

Figure~\ref{fig:figure_transition_stark_on} shows that, for each of the three perturbations, the defect state has the smallest \(N_{\mathrm{ch}}\) among the displayed initial eigenstates.
For carrier modulation, its two channels are the first doublet identified in the main text.
The sideband perturbations connect the defect state to more levels, but still to fewer states than the finite-energy eigenstates shown here.
This is a statement about first-order connectivity at the chosen operating point; it does not preclude sequential transitions at finite drive strength.

\subsection{Response of finite-energy eigenstates}
\label{subsec:eigenstate_response}

The spin observable in Eq.~\eqref{eqn:defect_spin_response} is well matched to the spin-polarized defect state, but it can have little contrast for investigating transitions between finite-energy eigenstates with similar spin populations.
For the latter states we instead use their survival probability
\begin{equation}
P_i(t)=\operatorname{Tr}
\left[\rho(t)\ket{\psi_i}\!\bra{\psi_i}\right]
\label{eqn:survival_probability}
\end{equation}
and define
\begin{equation}
D_i=
\sqrt{\left\langle
\left[P_i(t)_{\mathrm{mod}}-P_i(t)_0\right]^2
\right\rangle_{1\le f_Nt\le4}}.
\label{eqn:survival_response}
\end{equation}
The simulations are initialized in the corresponding exact static eigenstate.
We calculate \(D_{1,+}\) and \(D_{2,+}\) for the two regimes and the three modulation protocols in Fig.~\ref{fig:figure_eigenstate_response}.
Since \(D_i\) and \(\Delta_0\) refer to different observables, their numerical magnitudes cannot be compared directly.

The map colors use one common \(0\le D_i\le1\) scale with a power-law normalization of exponent \(1/2\), which increases the visibility of weak features.
Some first-order allowed transitions remain weak since their coupling matrix elements are small or the finite observation time limits population transfer.
Conversely, strong modulation can populate states through multiple steps.
The coupling matrices in Fig.~\ref{fig:figure_transition_stark_on} show which channels are present at first order.

\subsection{Stark-shift-free comparison}
\label{subsec:stark_free}

We repeat the coupling and response calculations after setting \(\delta_S=0\).
The carrier and phase-modulation operators retain the forms in Eq.~\eqref{eqn:modulation_operators}.
For amplitude modulation, however, removing the Stark term also removes \((\partial\delta_S/\partial g_{\min})\hat\sigma_z\) from \(\hat O_g\).
The Stark-free comparison therefore changes both the energy gaps and, for this protocol, the perturbation operator itself.

Figure~\ref{fig:figure_transition_stark_off} uses exactly the same state ranges and the same absolute coupling scale as Fig.~\ref{fig:figure_transition_stark_on}.
The defect state again has the fewest direct channels for every perturbation among the displayed states.
The changes in the sideband-amplitude coupling matrix arise from the altered eigenstate mixing together with the removal of the Stark-shift derivative from the perturbation operator.

Figure~\ref{fig:figure_response_stark_off} gives the corresponding finite-drive response for \(\ket{\psi_0}\), \(\ket{\psi_{1,+}}\), and \(\ket{\psi_{2,+}}\).
To compare all three initial states with a common observable, this figure uses the survival response \(D_i\) in Eq.~\eqref{eqn:survival_response}, including \(D_0\) for the defect state.
The fixed Stark-free spectrum is shown next to each response map, with the transitions allowed by the corresponding modulation operator highlighted.
In the Stark-free model, spin-sublattice exchange maps the amplitude-modulation responses of the two coupling regimes onto each other.
The same correspondence holds for weaker-sideband frequency modulation when the phase excursion is also reversed.
We therefore show the \(g_1>g_2\) regime as a representative case.

\section{Phase-space tomography of the winding trajectories}
\label{sec:phase_space}

\subsection{Conditional motional-state fidelity}
\label{subsec:motional_fidelity}

Characteristic-function tomography is performed after postselecting the spin state on the occupied defect-state sublattice.
Let \(\rho_{\mathrm{mot}}\) be the normalized conditional motional state and
\begin{equation}
\mathcal X(\beta)=
\operatorname{Tr}\!\left[\rho_{\mathrm{mot}}\hat D(\beta)\right]
\label{eqn:characteristic_function}
\end{equation}
its symmetrically ordered characteristic function.
For the pure ideal motional state \(\rho_{\mathrm{ideal}}\), the overlap identity gives
\begin{align}
F_{\mathrm{mot}}
&=\operatorname{Tr}(\rho_{\mathrm{mot}}\rho_{\mathrm{ideal}})\nonumber\\
&=\frac{1}{\pi}\int d^2\beta\,
\mathcal X_{\mathrm{exp}}^*(\beta)
\mathcal X_{\mathrm{ideal}}(\beta).
\label{eqn:characteristic_overlap}
\end{align}
The integral is evaluated on the measured grid.
Let \(q_{\mu j}\) denote the raw bright-state fraction for the real (\(\mu=\mathrm R\)) or imaginary (\(\mu=\mathrm I\)) component at grid point \(\beta_j\), estimated from \(N=50\) detection shots.
Propagating the binomial statistical uncertainty through the linear detection-contrast correction and the conversion from population to characteristic function gives
\begin{equation}
\sigma_{\mu j} = \frac{2}{C} \sqrt{\frac{q_{\mu j}(1-q_{\mu j})}{N}}, \qquad C=0.92,
\label{eqn:characteristic_uncertainty}
\end{equation}
where \(\sigma_{\mathrm Rj}\) and \(\sigma_{\mathrm Ij}\) are the statistical uncertainties of the real and imaginary components, respectively.

For the uniform measurement grid, let \(\Delta A=\Delta\beta_{\mathrm R}\Delta\beta_{\mathrm I}\) be the area associated with each grid point.
Assuming independent measurements, the propagated fidelity variance is
\begin{equation}
\sigma_{F_{\mathrm{mot}}}^{2} = \left(\frac{\Delta A}{\pi}\right)^2 \sum_j
\left[ \left(\operatorname{Re}\mathcal X_{\mathrm{ideal}}(\beta_j)\right)^2 \sigma_{\mathrm Rj}^{2} + \left(\operatorname{Im}\mathcal X_{\mathrm{ideal}}(\beta_j)\right)^2 \sigma_{\mathrm Ij}^{2}\right].
\label{eqn:fidelity_uncertainty}
\end{equation}
The detection contrast and the calibration parameters entering \(\mathcal X_{\mathrm{ideal}}\) are treated as fixed.

For each coupling regime and value of \(r_c\), the mean fidelity over the \(N_\phi=13\) measured phase settings and its propagated statistical uncertainty are
\begin{equation}
\bar F=\frac{1}{N_\phi}\sum_{k=1}^{N_\phi}F_k, \qquad
\sigma_{\bar F} = \frac{1}{N_\phi} \sqrt{\sum_{k=1}^{N_\phi}\sigma_{F_k}^{2}},
\label{eqn:mean_fidelity_uncertainty}
\end{equation}
where \(F_k\) is the conditional motional-state fidelity at the \(k\)th phase setting.
Measurements at different phase settings are assumed to be independent. 
All reported fidelity uncertainties correspond to one standard deviation.

\begin{figure}[t!]
\centering
\includegraphics[width=0.9\linewidth]{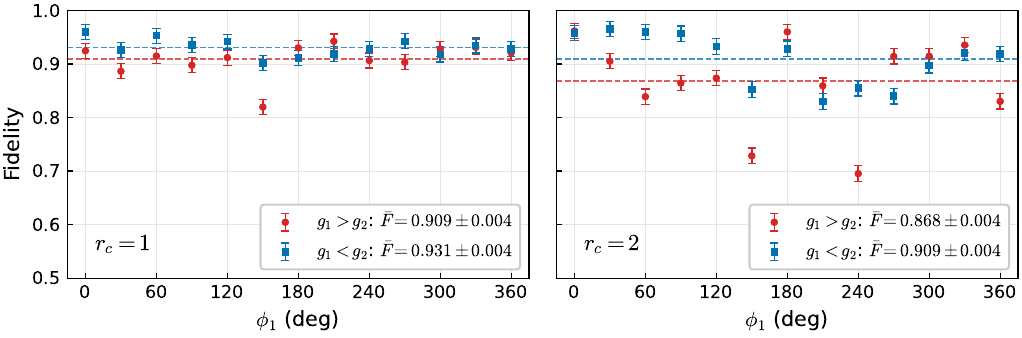}
\caption{
Conditional motional-state fidelity from the characteristic-function overlap.
The left and right panels show \(r_c=1\) and \(r_c=2\), respectively, at \(r_s=0.25\) and \(\phi_2=0\).
Red circles denote \(g_1>g_2\), and blue squares denote \(g_1<g_2\).
Error bars denote one-standard-deviation statistical uncertainties propagated from the characteristic-function measurements. Dashed lines indicate the mean fidelity over the measured phase settings for each coupling regime and value of \(r_c\). The legends report these means and their propagated one-standard-deviation statistical uncertainties.
}
\label{fig:figure_motional_fidelity}
\end{figure}

Figure~\ref{fig:figure_motional_fidelity} summarizes the results for all phases used to construct the winding trajectories. For \(r_c=1\), the mean fidelities over the measured phase settings are \(0.909\pm0.004\) and \(0.931\pm0.004\) for \(g_1>g_2\) and \(g_1<g_2\), respectively. For \(r_c=2\), the corresponding means are \(0.868\pm0.004\) and \(0.909\pm0.004\). The \(r_c=2\) fidelities exhibit a stronger phase dependence, while the representative point displayed in Fig.~\ref*{main-fig:figure2} of the main text has \(F_{\mathrm{mot}}=0.915\pm0.014\).

\subsection{Measured characteristic functions}
\label{subsec:characteristic_functions}

Figures~\ref{fig:figure_char_rc1_first}--\ref{fig:figure_char_rc2_second} show the complete characteristic-function data used for the trajectory analysis.
The parameters are \(r_s=g_{\min}/g_{\max}=0.25\) and \(\phi_2=0\), while \(\phi_1\) is stepped through one full cycle.
For each phase and coupling regime, the figures compare the measured real and imaginary parts of \(\mathcal X(\beta)\) with the ideal characteristic function calculated from Eq.~\textup{(\ref*{main-eqn:eqn2})}.
Figures~\ref{fig:figure_char_rc1_first} and \ref{fig:figure_char_rc1_second} contain the \(r_c=1\) data, and Figs.~\ref{fig:figure_char_rc2_first} and \ref{fig:figure_char_rc2_second} contain the \(r_c=2\) data.

\begin{figure*}[p]
\centering
\includegraphics[width=\textwidth]{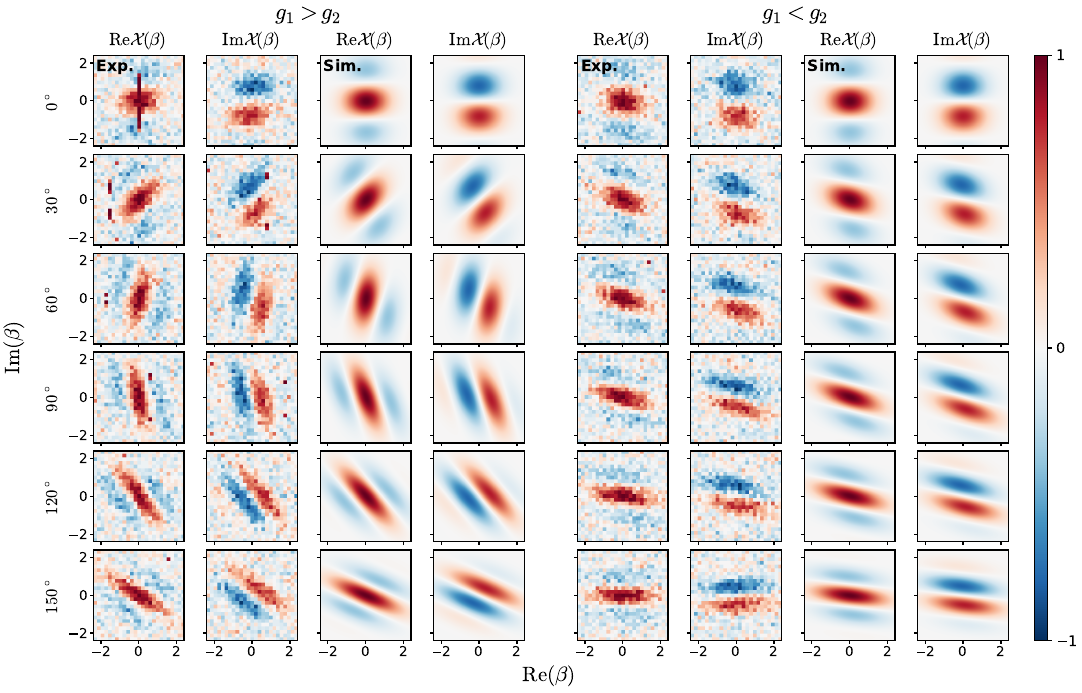}
\caption{
Measured (Exp.) and ideal (Sim.) characteristic functions for \(r_c=1\), \(r_s=0.25\), and \(\phi_2=0\), with \(\phi_1=0^\circ,30^\circ,\ldots,150^\circ\).
The left and right blocks correspond to \(g_1>g_2\) and \(g_1<g_2\), respectively.
Within each block, columns show \(\operatorname{Re}\mathcal X(\beta)\) and \(\operatorname{Im}\mathcal X(\beta)\) for experiment and the ideal defect state.
All panels use the common range \([-1,1]\).
}
\label{fig:figure_char_rc1_first}
\end{figure*}

\begin{figure*}[p]
\centering
\includegraphics[width=\textwidth]{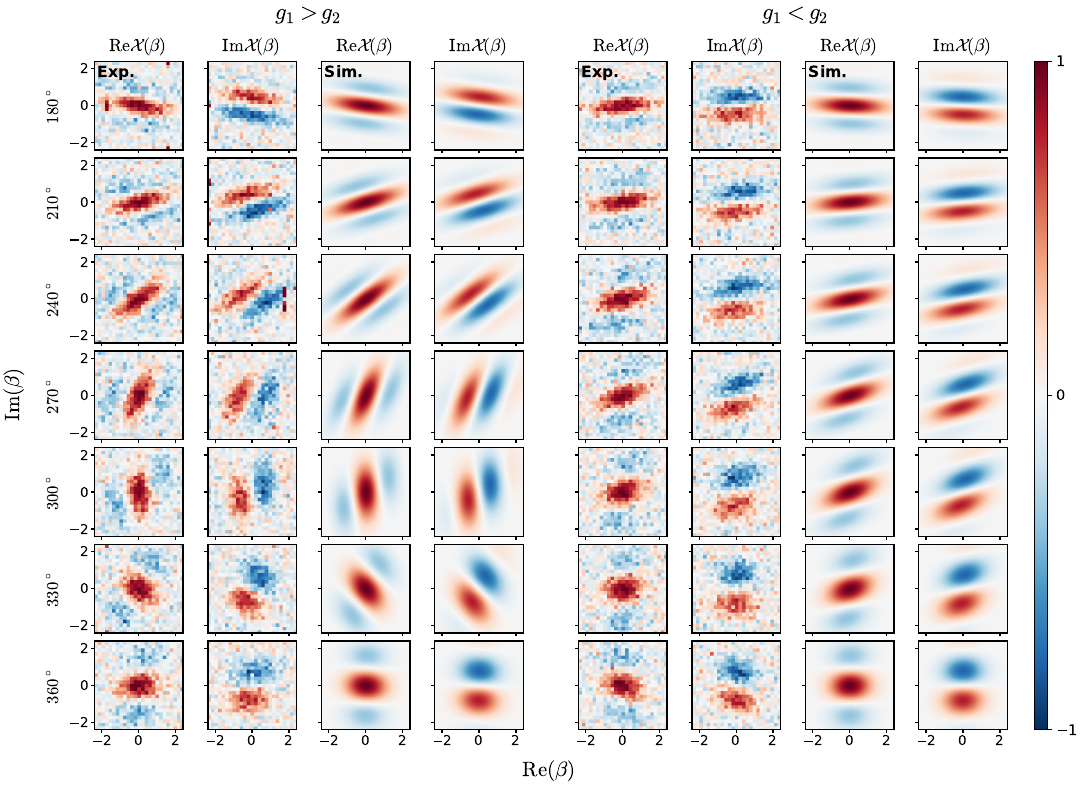}
\caption{
Continuation of Fig.~\ref{fig:figure_char_rc1_first} for \(r_c=1\) and \(\phi_1=180^\circ,210^\circ,\ldots,360^\circ\).
The point at \(360^\circ\) closes the phase cycle and repeats the Hamiltonian at \(0^\circ\).
}
\label{fig:figure_char_rc1_second}
\end{figure*}

\begin{figure*}[p]
\centering
\includegraphics[width=\textwidth]{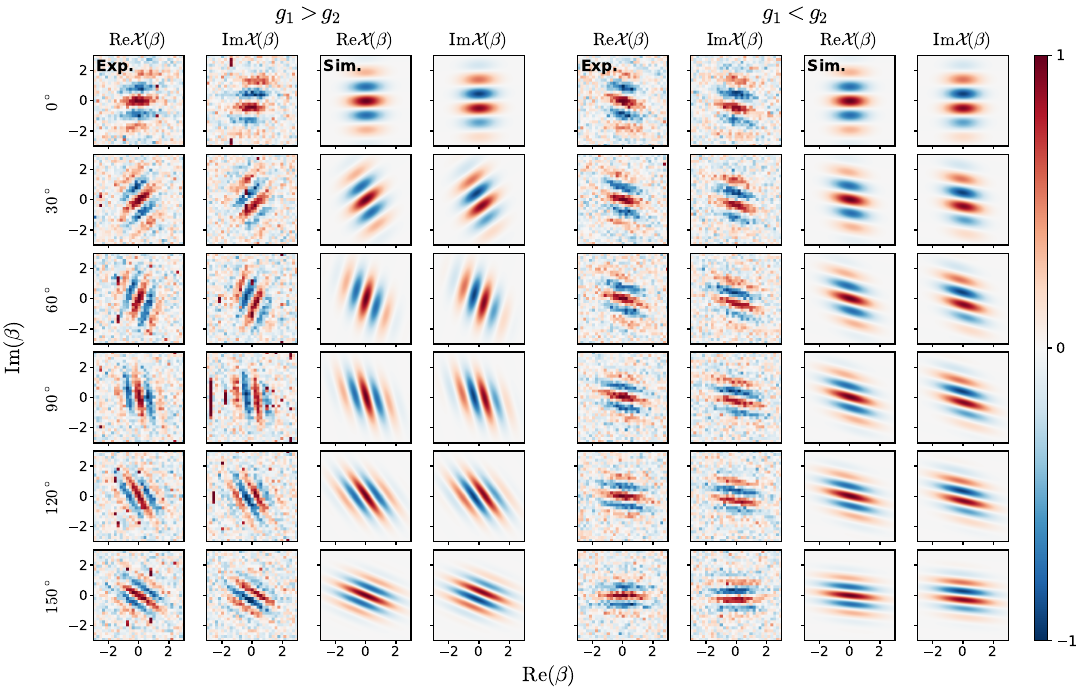}
\caption{
Measured (Exp.) and ideal (Sim.) characteristic functions for \(r_c=2\), \(r_s=0.25\), and \(\phi_2=0\), with \(\phi_1=0^\circ,30^\circ,\ldots,150^\circ\).
The arrangement and color range are the same as in Fig.~\ref{fig:figure_char_rc1_first}.
}
\label{fig:figure_char_rc2_first}
\end{figure*}

\begin{figure*}[p]
\centering
\includegraphics[width=\textwidth]{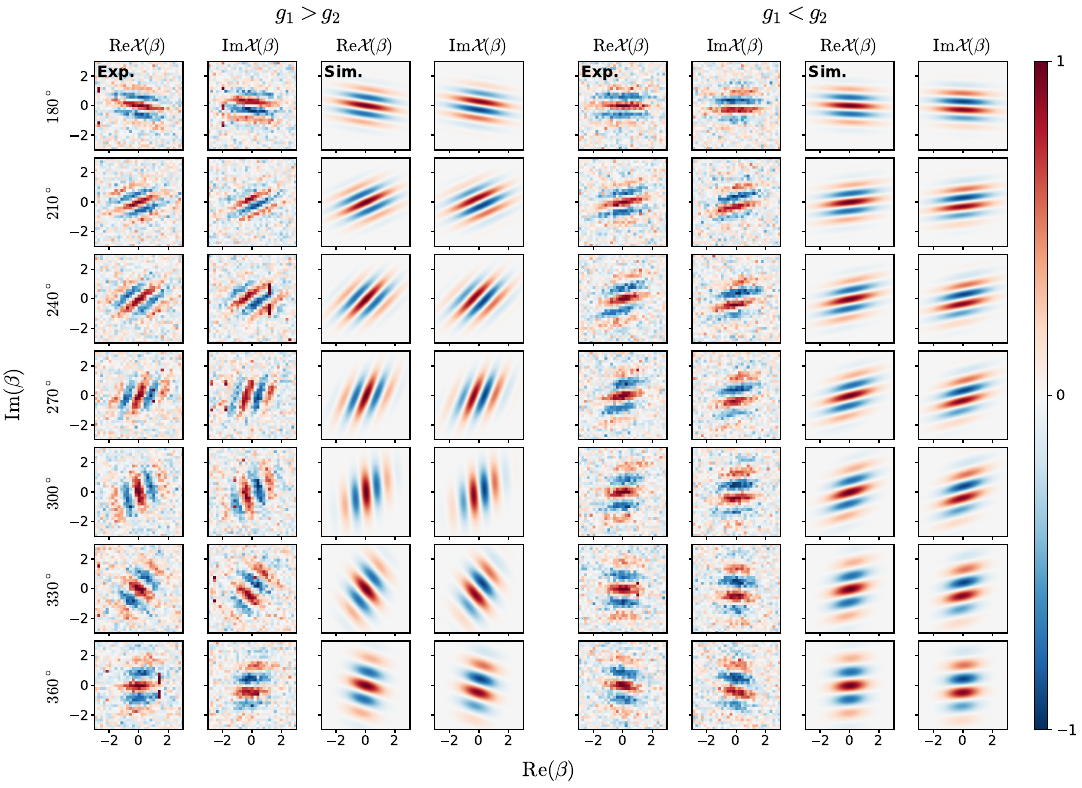}
\caption{
Continuation of Fig.~\ref{fig:figure_char_rc2_first} for \(r_c=2\) and \(\phi_1=180^\circ,210^\circ,\ldots,360^\circ\).
}
\label{fig:figure_char_rc2_second}
\end{figure*}

\subsection{Reconstructed Wigner functions and fitted centroids}
\label{subsec:wigner_functions}

We reconstruct the Wigner function from the measured characteristic function according to
\begin{equation}
\mathcal W(\gamma)=\frac{1}{\pi^2}
\int d^2\beta\,
\mathcal X(\beta)
e^{\gamma\beta^*-\gamma^*\beta}.
\label{eqn:wigner_reconstruction}
\end{equation}
To determine the centroids, we fit each reconstructed Wigner distribution to the Wigner function of the displaced-squeezed defect state by minimizing the sum of squared differences on the phase-space grid.
The carrier amplitude is held at its nominal value, while the two sideband amplitudes and phases are free fit parameters.
The fitted parameters define \(\mathcal W_{\mathrm{fit}}\) and give the centroids used in Fig.~\ref*{main-fig:figure3} through Eq.~\textup{(\ref*{main-eqn:eqn3})}.
In addition to the fixed offsets applied during state preparation (Sec.~\ref{sec:phase_calibration}), the analysis accounts for residual red-sideband phase offsets determined from the independent calibration measurements at each phase setting.
The ideal Wigner functions \(\mathcal W_{\mathrm{ideal}}\) are calculated from Eq.~\textup{(\ref*{main-eqn:defect_parameters})} using the nominal coupling amplitudes and the corrected red-sideband phase, with \(\phi_2=0\).
The same phase correction is used for the ideal characteristic functions and the overlap fidelities.
The phase labels in the figures denote the nominal settings.

\begin{figure*}[p]
\centering
\includegraphics[width=\textwidth]{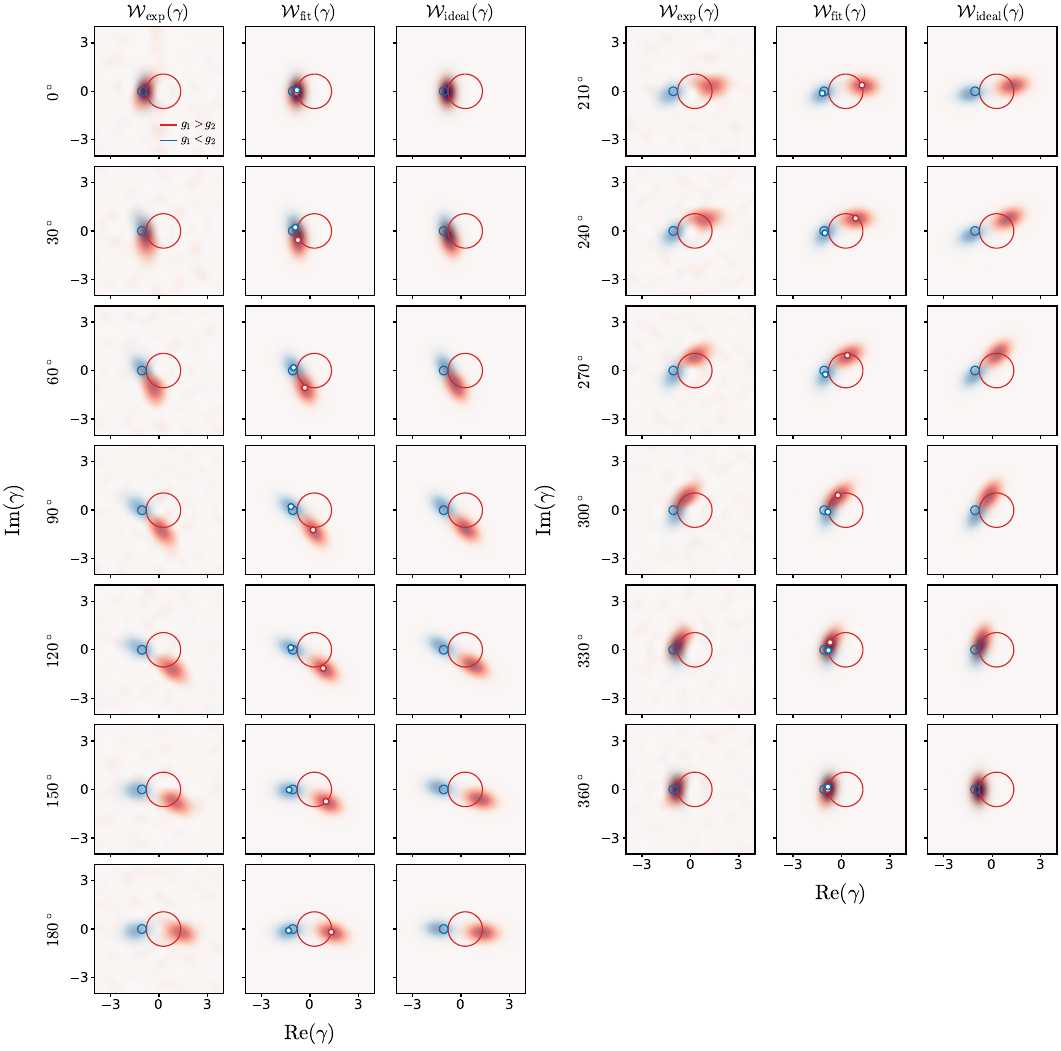}
\caption{
Wigner functions for \(r_c=1\), \(r_s=0.25\), and \(\phi_2=0\), for all sampled values of \(\phi_1\).
Columns show the experimental reconstruction \(\mathcal W_{\mathrm{exp}}\), fitted defect-state distribution \(\mathcal W_{\mathrm{fit}}\), and ideal distribution \(\mathcal W_{\mathrm{ideal}}\).
Red and blue overlays correspond to \(g_1>g_2\) and \(g_1<g_2\), respectively; the curves show the predicted trajectories.
White-centered red and blue markers in the fitted panels show the extracted centroids used in Fig.~\ref*{main-fig:figure3}.
}
\label{fig:figure_wigner_rc1}
\end{figure*}

\begin{figure*}[p]
\centering
\includegraphics[width=\textwidth]{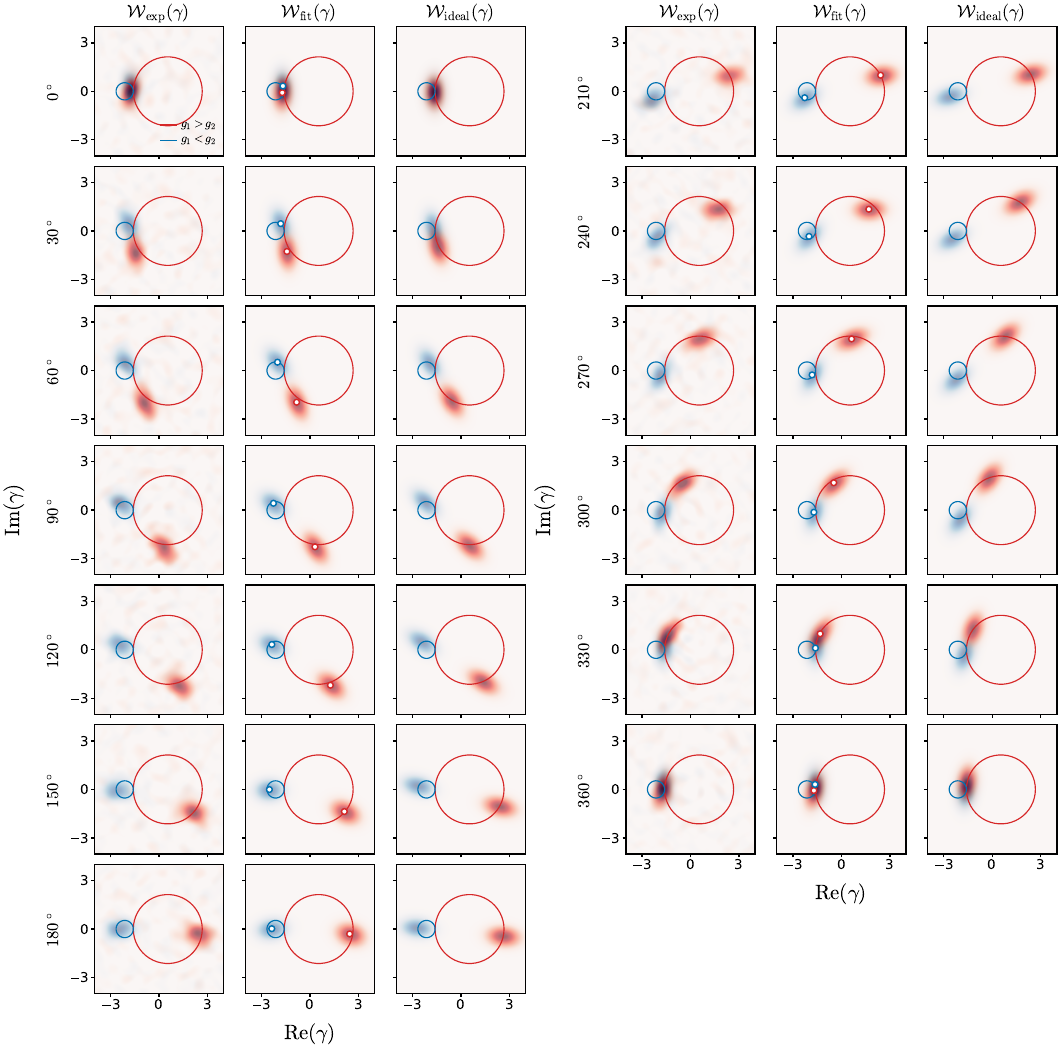}
\caption{
Wigner functions for \(r_c=2\), \(r_s=0.25\), and \(\phi_2=0\), with the same arrangement and conventions as in Fig.~\ref{fig:figure_wigner_rc1}.
}
\label{fig:figure_wigner_rc2}
\end{figure*}

Figures~\ref{fig:figure_wigner_rc1} and \ref{fig:figure_wigner_rc2} overlay the two coupling regimes in each panel: red represents \(g_1>g_2\), and blue represents \(g_1<g_2\).
These colors label two separately reconstructed Wigner functions and do not represent positive and negative quasiprobability.
The red and blue curves are the predicted centroid trajectories from Eq.~\textup{(\ref*{main-eqn:eqn3})}.
The white-centered markers in the fitted panels are the extracted centroids.
The experimental, fitted, and ideal overlays use the same range, \(0\le\mathcal W\le2/\pi\), for each regime.

\enlargethispage{2\baselineskip}
\renewcommand{\bibsection}{\section*{References}}
\makeatletter
\def\@bibstyle{main}
\makeatother
\bibliography{main}